\documentclass[%
 aip,
 jap,%
 amsmath,amssymb,
reprint,%
floatfix
]{revtex4-1}

\draft

\usepackage{graphicx}% Include figure files
\usepackage{sidecap}

\usepackage{dcolumn}% Align table columns on decimal point
\usepackage{chemfig} 
\usepackage{isotope}
\usepackage{braket}

\usepackage[version=3]{mhchem}
\usepackage{miller}

\usepackage{siunitx} % to get proper distances between num and unit 
\usepackage{xspace}
\usepackage{xargs} % Use more than one optional parameter in a new commands
\usepackage{multirow}
\usepackage{nicefrac}
\usepackage[colorinlistoftodos,prependcaption,textsize=tiny]{todonotes}
\newcommandx{\unsure}[2][1=]{\todo[linecolor=red,backgroundcolor=red!25,bordercolor=red,#1]{#2}}
\newcommandx{\change}[2][1=]{\todo[linecolor=blue,backgroundcolor=blue!25,bordercolor=blue,#1]{#2}}
\newcommandx{\info}[2][1=]{\todo[linecolor=gray,backgroundcolor=gray!25,bordercolor=gray,#1]{#2}}
\newcommandx{\improvement}[2][1=]{\todo[linecolor=Plum,backgroundcolor=Plum!25,bordercolor=Plum,#1]{#2}}
\newcommandx{\thiswillnotshow}[2][1=]{\todo[disable,#1]{#2}}

\usepackage{hyperref}% add hypertext capabilities

\newcommand{\nv}{$\text{N}_\text{As}\text{V}_\text{Ga}$\xspace}
\newcommand{\CThreeV}{\textit{C}$_\text{3v}$\xspace}
\newcommand*\rb[1]{\textcolor{black}{#1}}
 
\makeatletter
\newcommand\xleftrightarrow[2][]{%
  \ext@arrow 9999{\longleftrightarrowfill@}{#1}{#2}}
\newcommand\longleftrightarrowfill@{%
  \arrowfill@\leftarrow\relbar\rightarrow}
\makeatother
\begin{document}

\preprint{Special Topic on Defects in Semiconductors}

\title{Piezospectroscopy and first-principles calculations of the nitrogen-vacancy center in gallium arsenide}% 

\author{Nicola Kovac}%
% \email{nkovac@hm.edu}
\author{Christopher K\"unneth}

%\altaffiliation[Also at ]{Physics Department, XYZ University.}%Lines break automatically or can be forced with \\

%\email{rmaterli@hm.edu}
%\author{Hans Edwin Wagner}
%\affiliation{%
    %Munich University of Applied Sciences, Department of Applied Sciences and Mechatronics, Lothstr. 34, D-80335 Munich, Germany
%}%

% \author{Alfred Kersch}%

\author{Hans Christian Alt}%
\email{hchalt@hm.edu}
\affiliation{%
    Munich University of Applied Sciences, Department of Applied Sciences and Mechatronics, Lothstr. 34, D-80335 Munich, Germany
}%

\date{\today}% It is always \today, today,
             %  but any date may be explicitly specified

\begin{abstract}
The nitrogen-vacancy (NV) center occurs in GaAs bulk crystals doped or implanted with nitrogen. The local vibration of nitrogen gives rise to a sharp infrared absorption band at \SI{638}{cm^{-1}}, exhibiting a fine structure due to the different masses of neighboring \isotope[69]{Ga} and \isotope[71]{Ga} host isotopes. Piezospectroscopic investigations in the crystallographic \hkl<100> direction prove that the center has \CThreeV point symmetry, which is weakly perturbed by the isotope effect. The stress-induced shifts of some band components  show an unusual non-linear behavior that can be explained by coupling between the isotope and the stress splitting. First-principles density-functional theory calculations are in full accordance with the experiments and confirm the \CThreeV symmetry, caused by relaxation of the nitrogen atom from the anion lattice site towards the nearest-neighbor Ga plane. \rb{Furthermore, the calculations indicate the $-3$ charge state of the center as the most stable one for nearly all Fermi level positions.} The NV center in GaAs is structurally analogous to the same center in diamond. 
\end{abstract}

%\pacs{Valid PACS appear here}% PACS, the Physics and Astronomy
                             % Classification Scheme.
 %Use showkeys class option if keyword
                              %display desired
\maketitle

\section{Introduction}
Nitrogen is of continuing interest as a dopant in GaAs. Renewed research activities have been initiated due to the development of dilute nitrides, in particular of the alloys GaAs$_{1-x}$N$_x$ and Ga$_{1-y}$In$_{y}$As$_{1-x}$N$_{x}$, with N fractions up to a few percent.\cite{Weyers1992} The incorporation of nitrogen on the anion lattice site (N$_\text{As}$) leads to a strong narrowing of the band gap of initially \SI{180}{meV} per percent of nitrogen. There is a great technological interest in these alloys, driven by the search for optimized materials for lasers emitting at 1.3 and \SI{1.55}{\mu \meter} as well as multi-junction solar cells.\cite{Kondow1997,Jackrel2007}

The importance of nitrogen as a luminescent center in GaAs has been first put forward in 1985, when luminescence from isoelectronic N and NN pairs was detected in N doped GaAs layers under pressure\cite{Wolford1985}, and later by more extensive macro and micro photoluminescence studies.\cite{Liu1990,Francoeur2004} Recently, a nitrogen-related center acting as a single-photon emitter has been reported from nitrogen delta-doped layers in GaAs. These centers might \rb{be} of interest for future quantum information processing.\cite{Ikezawa2012,Ikezawa2017}

These materials issues of N in the GaAs lattice are primarily concerned with isoelectronic substitutional nitrogen. Consequently, detailed theoretical investigations of the formation of isolated N, different NN pairs and N clusters have been carried out.\cite{Kent2001} However, interstitial N, complexes of N with Ga and As atoms and with intrinsic defects (interstitials, vacancies) also play an important role in these applications, as evidenced both theoretically and experimentally.\cite{Janotti2003,Spruytte2001,Toivonen2003}

Nitrogen-related defects in GaAs can be investigated by local vibrational mode (LVM) spectroscopy, due to the small mass of the N atom compared to the Ga and As atoms of the lattice. The isolated substitutional nitrogen atom on anion site, N$_\text{As}$, gives rise to a relatively broad unstructured LVM band at \SI{471.5}{cm^{-1}} (low temperatures).\cite{Kachare1973,Alt1997} 

Another LVM band at \SI{638}{cm^{-1}}, first tentatively associated with nitrogen impurities due to its appearance in nitrogen-rich crystals \cite{Gaertner1999}, is caused by the N$_\text{As}$-V$_\text{Ga}$ nearest-neighbor pair, the nitrogen-vacancy (NV) center in GaAs.\cite{Alt2004} It occurs in both GaAs bulk material doped with a high concentration of N and in N-implanted GaAs layers. The LVM at \SI{638}{cm^{-1}} splits at low temperature into four components due to the mass effect of the different host isotopes \isotope[69]{Ga} and \isotope[71]{Ga}, acting as nearest neighbors of the substitutional nitrogen atom. An empirical valence-force model was given which reproduces the fine structure of the \SI{638}{cm^{-1}}-LVM both qualitatively and quantitatively.\cite{Alt2004} The analysis is based on a twofold degenerate \textit{E} (transverse) mode of a trigonal center with \CThreeV symmetry, weakly perturbed by the different Ga host isotope masses. Finally, it should be mentioned that the related nitrogen-hydrogen-vacancy center, incorporating additionally one hydrogen atom, was also investigated by LVM spectroscopy.\cite{Ulrici2005}

In this work an extensive study on the atomistic structure, energetics and vibrational properties of the NV center in GaAs is presented. We performed FTIR absorption measurements on the \SI{638}{cm^{-1}}-band under uniaxial stress in \hkl<100> direction to get further experimental information on the defect symmetry. Secondly, density-functional theory (DFT) calculations of the minimum energy configurations in different charge states were carried out to calculate the formation energy and the associated LVM frequencies.

\section{Experimental and computational methods}

Samples investigated were prepared from a nitrogen-doped GaAs crystal grown by the liquid-encapsulation Czochralski technique. Nitrogen doping was achieved by applying a N$_2$ gas atmosphere of \SI{70}{\bar} during crystal growth. For uniaxial stress experiments, rod-like specimens with a typical size of \SI{2x4x10}{mm} were cut with alignment of the long axis parallel to the crystallographic \hkl[001] direction. Uniaxial stress measurements up to \SI{0.15}{GPa} were performed using a home-made push-rod apparatus inside an optical helium cryostat. The force on the mounted sample was provided by a pressure intensifier coupled to the apparatus. The absolute value of stress was set by the nitrogen gas pressure on the low-pressure side of the pressure intensifier, which was monitored by a precision gauge. Fourier transform infrared (FTIR) measurements were performed with a vacuum instrument (Bruker Vertex 80v), equipped with a potassium bromide (KBr) beam splitter and a liquid-nitrogen cooled mercury cadmium telluride (MCT) detector. The absorption spectra were taken with an apodized spectral resolution of \SI{0.08}{cm^{-1}}. A wire-grid polarizer in front of the cryostat was used to obtain polarized spectra.

The minimum-energy structures were obtained with the all-electron DFT code FHI-Aims\cite{Blum2009,Knuth2015,Marek2014,Auckenthaler2011,Havu2009} which utilizes numeric atom-centered basis functions. The tight basis set on the "Second tier"-level in the LDA approximation (PW\cite{Perdew1992} parameterization) was used for all calculations in this publication. A pure GaAs super cell with 64 atoms $\text{Ga}_{32} \text{As}_{32}$ and F$\overline{4}3$m (No. 216) symmetry was set up as the host crystal for the \nv (NV) center (63 atoms for NV, $\text{Ga}_{31} \text{As}_{31} \text{N}$). A k-point grid of \num{3x3x3} was used and the atomic positions and lattice constants were relaxed until the forces fell below \SI{5e-5}{eV/\AA} and \SI{5e-4}{eV/\AA} for the electronic and ionic forces, respectively. Only in the case of charged ($q \neq 0$) super cells, the lattice constants were kept fixed at the uncharged values, relaxing only the atomic positions. 

Formation energies were calculated according to \cite{Freysoldt2014}
\begin{eqnarray}
E^\text{f}&& \left[\text{NV}^q\right] = E_\text{tot}\left[\text{NV}^q\right] - E_\text{tot}\left[\text{GaAs}\right] \nonumber\\ 
&& + \mu_{\text{N}} - \mu_\text{As} - \mu_\text{Ga}  \nonumber\\ 
&&  + q\left( E_\text{F} + E_\text{VBM}\left[\text{GaAs}\right] + \Delta V\left[\text{NV}^0\right] \right) \nonumber\\
&&+ E_\text{corr}[\text{NV}^q]
\label{eq:formation_energy}
\end{eqnarray}
with $E^\text{f}$ the formation energy, $E_\text{tot}$ the total energies, $\mu$ the chemical potentials, $E_\text{F}$ the Fermi level referenced to the energy of the valence band maximum $E_\text{VBM}$, $\Delta V$ the potential alignment, and $E_\text{corr}$ the charge correction due to finite size of the unit cell. $E_\text{corr}$ was calculated with the scaling law \cite{MakovPayne1995} $E_\text{f} \sim \nicefrac{a}{L} + c$ using 216 atomic supercells. $a$ and $c$ are fit parameters and $L$ is the size of the super cell. Charged structures were considered for the charges $q/e=-4,\ldots,+4$, with the lattice constants fixed to the uncharged structure. $\mu_\text{As}$ was calculated from trigonal metallic As, $\mu_\text{Ga}$ from metallic $\alpha$-Ga and $\mu_{\text{N}}$ from the N$_2$ molecule. 

Figures of the atomic structures in this publication are produced with Ovito.\cite{Stukowski2010} Vibrational frequencies were obtained with the utility Phonopy \cite{Togo2015} using finite displacements. Isotope-dependent vibrational frequencies were calculated by changing the masses in the denominator of the dynamical matrix before solving the eigenvalue problem. The calculated vibrational frequencies ($\omega_\text{calc}$) were scaled to the experimental frequencies ($\omega_\text{exp}$) with the scaling factor $c = \left(\sum_i \omega_{i,\text{calc}} \omega_\text{i,\text{exp}}\right) / \sum_i \omega_{i,\text{calc}}^2 $ according to Ref. \onlinecite{Irikura2005}.

\section{Results and discussion}
\subsection{Piezospectroscopic experiments}

\begin{figure}
\includegraphics[width=0.50\textwidth]{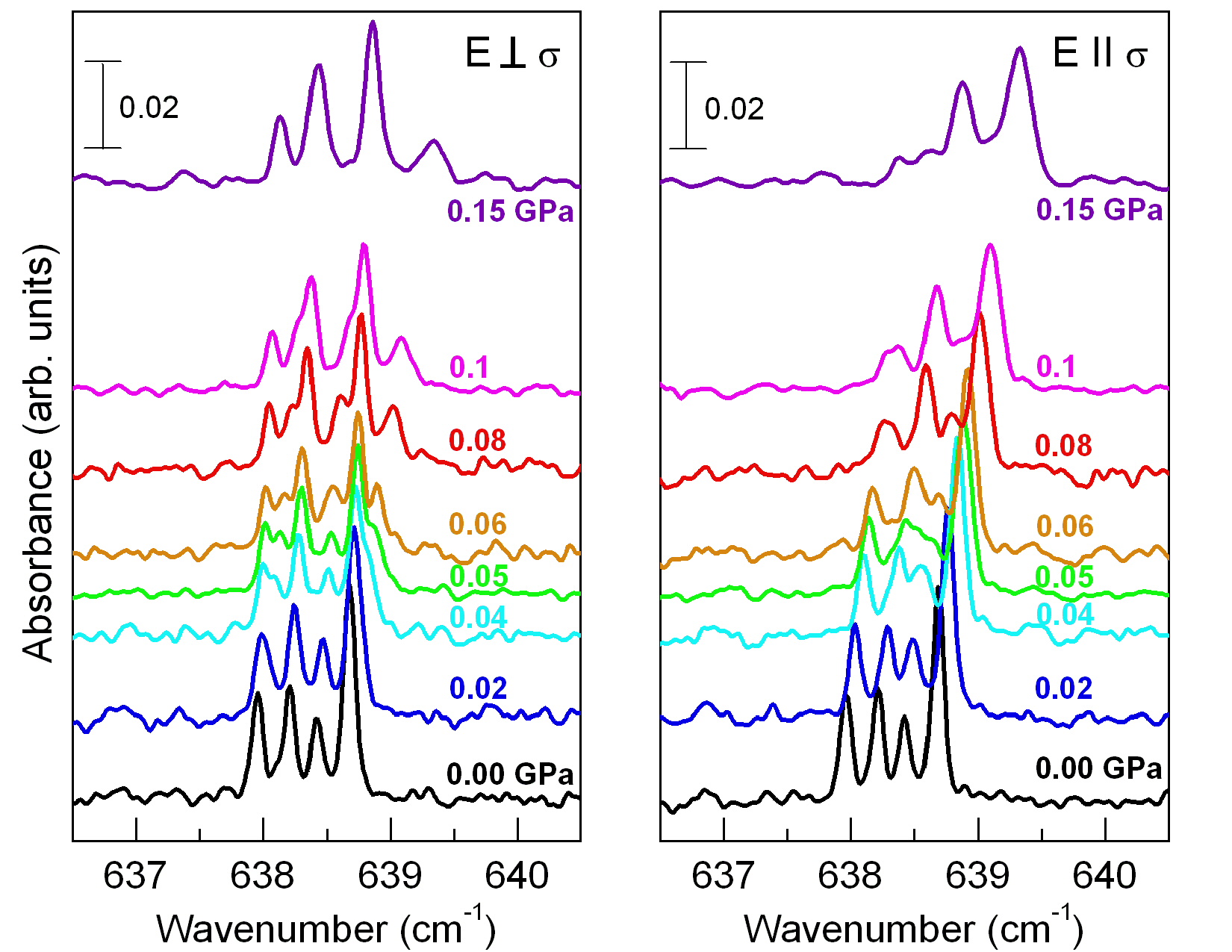}
\caption{\label{fig:spectra_under_stress} Effect of \hkl<100> stress on the fine structure of the \SI{638}{cm^{-1}}-band for both polarization directions. Sample temperature is \SI{10}{K} and the spectral resolution \SI{0.08}{cm^{-1}}.}
\end{figure}

\begin{figure}
\includegraphics[width=0.5\textwidth]{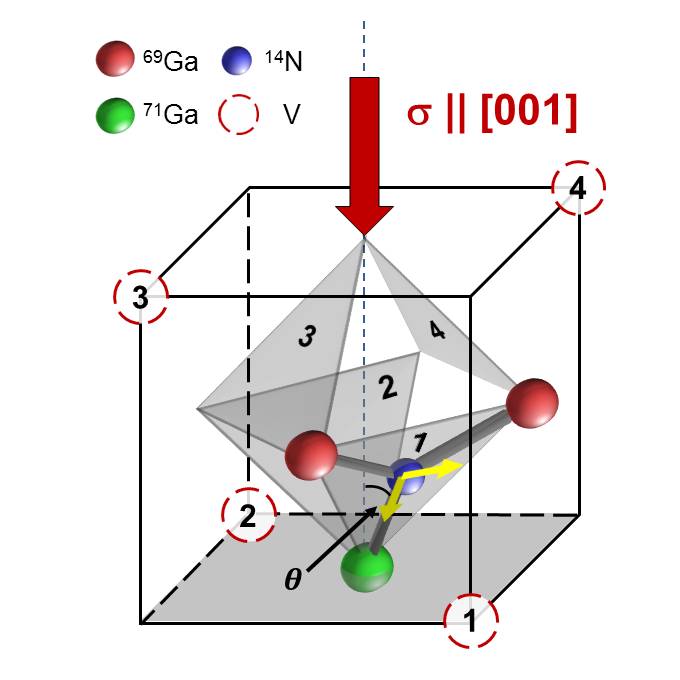}
\caption{\label{fig:orientation_model} Schematic picture of the geometry and orientations of the NV complex (assumed to be planar, \rb{see text for discussion}) in the GaAs lattice. The possible vacancy positions are labelled by 1, 2, 3, and 4. The Ga plane forms an angle $\theta$ with the \hkl[001] stress direction. In isotopically mixed configurations (as shown here), the different possible orientations of the eigenvector (\rb{yellow arrows}) for the\textit{A'} and \textit{A''} mode have to be taken into account for IR intensity calculations.}
\end{figure}

% Raw frequencies from DFT
% om_dft = np.array([646.85, 646.86, 647.12, 647.63, ]) 
% om_exp = np.array([637.95,  638.41,  638.20,  638.67, ])

% before scaling 
% 646.85, 646.86, 647.36, 647.12,  647.61, 647.63,
% After scaling with c
% 638.05 638.06 638.55 638.31 638.80 638.82 

\begin{table*}
\caption{\label{tab:uniaxial_stress}  NV center/\SI{638}{cm^{-1}}-band: Isotopic combinations and point symmetries, LVM symmetry types, weighted probabilities, frequencies, stress-induced splitting and polarization behavior. The calculated vibrational frequencies are for charge $q/e=-3$ and scaled with a factor of $c = \num{0.9862}$.}
\begin{ruledtabular}
\begin{tabular}{cccccccc}
Configuration & Point & LVM symm. type  & Weighted  & Frequency/\si{cm^{-1}}  & Frequency shift  &  \multicolumn{2}{c}{Intensity ratio} \\
 (Notation) & symmetry &  (Degeneracy)  &  probability\footnotemark[1]  &  calc. (meas.)  & \hkl<100> stress\footnotemark[2] & I$_\parallel : \text{I}_\perp$ & I$_\parallel : \text{I}_\perp$\footnotemark[3] \\
\hline
\isotope[14]{N}\isotope[71]{Ga}$_3$ &  \multirow{2}{*}{\CThreeV} & \multirow{2}{*}{\textit{E} (2)} & \multirow{2}{*}{0.0635}  &   \multirow{2}{*}{ 637.93 (637.95)} & ($A_\alpha - 2B_\alpha$)$\sigma$ & $4:1$ &  \\
 ($\alpha$) & & & & & ($A_\alpha + 2B_\alpha$)$\sigma$ & $0:3$ & \\\hline
 
  &  \multirow{4}{*}{\textit{C}$_{\text{s}}$} & \multirow{2}{*}{\textit{A'} (1)} &  \multirow{2}{*}{0.1435}  & \multirow{2}{*}{ 638.43 (638.41)} & $A_{\beta 1}\sigma$ & $\nicefrac{8}{3} : \nicefrac{2}{3}$ & \\
\isotope[14]{N}\isotope[71]{Ga}$_2$\isotope[69]{Ga}$_1$ & & & & & $A_{\beta 2}\sigma$  &  $\nicefrac{4}{3} : \nicefrac{10}{3}$ &  $\nicefrac{16}{3} : \nicefrac{4}{3}$ \\
($\beta$) & & \multirow{2}{*}{\textit{A''} (1)} & \multirow{2}{*}{0.1435} &  \multirow{2}{*}{ 637.94 (637.95) } & $A_{\beta 3}\sigma$  &  $0 : 2$ & \\
& & & & & $A_{\beta 4}\sigma$  &  $4:2$ &  $0 : 4$  \\\hline
 
  & \multirow{4}{*}{\textit{C}$_{\text{s}}$} & \multirow{2}{*}{\textit{A'} (1)} &  \multirow{2}{*}{0.2162} &  \multirow{2}{*}{ 638.19 (638.20)} & $A_{\gamma 1} \sigma$ & $\nicefrac{8}{3} : \nicefrac{2}{3}$ &  \\
\isotope[14]{N}\isotope[71]{Ga}$_1$\isotope[69]{Ga}$_2$& & & & & $A_{\gamma 2} \sigma$  &  $\nicefrac{4}{3} : \nicefrac{10}{3}$ & $\nicefrac{16}{3} : \nicefrac{4}{3}$ \\
 ($\gamma$)  & & \multirow{2}{*}{\textit{A''} (1)} & \multirow{2}{*}{0.2162} &  \multirow{2}{*}{ 638.67 (638.67) } & $A_{\gamma 3} \sigma$  &  $0 : 2$ & \\
& & & & & $A_{\gamma 4} \sigma$  &  $4:2$ &  $0 : 4$  \\\hline

\isotope[14]{N}\isotope[69]{Ga}$_3$ &  \multirow{2}{*}{\CThreeV} & \multirow{2}{*}{\textit{E} (2)} & \multirow{2}{*}{0.2172} & \multirow{2}{*}{ 638.69  (638.67)} & ($A_\delta - 2B_\delta$)$\sigma$ & $4:1$ & \\
 ($\delta$)& & & & & ($A_\delta + 2B_\delta$)$\sigma$ & $0:3$ & \\
\end{tabular}
\end{ruledtabular}
    \footnotetext[1]{Calculated with the natural abundances of \isotope[69]{Ga} and \isotope[71]{Ga} of 0.61 and 0.39, respectively}
	\footnotetext[2]{\rb{Parameters $A$ and $B$ given in Table \ref{tab:parameters}}}
	\footnotetext[3]{Limiting case of complete stress-induced eigenvector rotation}
\end{table*}

Series of polarized absorption spectra of the \SI{638}{cm^{-1}}-band for different stresses $\sigma$ parallel to the \hkl[100] direction are displayed in FIG. \ref{fig:spectra_under_stress}. At zero stress the band has the known quadruplet fine structure originating from the isotopic perturbation of the LVM frequency by the different gallium host isotopes (\isotope[69]{Ga} and \isotope[71]{Ga}) as nearest neighbours of the nitrogen atom (\isotope[14]{N}). The corresponding isotopic configurations and LVM frequencies are listed in TAB. \ref{tab:uniaxial_stress}.The peak-to-peak separation of the four lines is about \SI{0.24}{cm^{-1}} and the full width at half-maximum (FWHM) amounts to \SI{0.11+-0.01}{cm^{-1}}. 

With increasing stress, all components of the band shift to higher frequencies. The individual line shifts are very small and lie between \num{0.1} and \SI{0.5}{cm^{-1}} per \SI{0.1}{GPa}. Moreover, some components show a splitting and others an apparent broadening with increasing stress. The line splitting is especially clearly seen in the series of perpendicularly polarized spectra at \rb{intermediate} stresses, whereas for parallel polarization this behavior is sometimes obscured by apparent broadening. However, the apparent line broadening as well as the simultaneous increase or decrease of line intensities can in fact be attributed completely to stress-induced splitting. As some lines still have a well-defined resolved line shape (e.g. at \SI{0.15}{GPa} for the (E $\perp$ $\sigma$) case), line broadening due to inhomogeneous stress fields inside the specimen occurs only to a minimal extent. From careful spectra analysis we estimate this increased width (FWHM) to about \SI{18}{\%} at highest stresses. Therefore, we assume that in both polarization directions the observed apparent line broadening results from overlapping of stress-split lines. Furthermore, comparison of the matching (\textit{E} $\perp$ $\sigma$) and (\textit{E} $\parallel$ $\sigma$) spectra shows that some split components are only partially polarized. Therefore, a clear visual separation and identification of the individual lines is not possible. That is why we started to evaluate the stress-induced splittings by curve-fitting methods, following the theoretical predictions of the piezospectroscopic behavior.

 The NV center has four different isotopic configurations due to the different combinations of the nearest-neighbor Ga atoms (TAB. \ref{tab:uniaxial_stress}) as well as four different orientations within the unit cell of the zinc-blende structure (FIG. \ref{fig:orientation_model}).\cite{Alt2004} Both isotopically pure configurations have \CThreeV symmetry and a twofold degenerate \textit{E} mode, whereas both mixed configurations have \textit{C}$_\text{s}$ symmetry and two non-degenerate modes of type \textit{A'} and \textit{A''}.  The characteristics of the uniaxial stress effect on non-degenerate states in cubic crystals have been given by Kaplyanskii.\cite{Kaply1964} The corresponding theory for doubly degenerate states in tetragonal and trigonal centers has been developed and applied to specific cases in Refs.  \onlinecite{Hughes1967,Davies1980,BechNielsen1989}. 
 
 For \hkl<100> stress, the \textit{A'} and \textit{A''} modes each split into two branches, due to partial lifting of orientational degeneracy.\cite{Kaply1964} The relative intensities for polarization parallel and perpendicular to the stress direction depend on the orientation of the dipole vector, given by the angle $\theta$' between this vector and the \hkl<100> stress direction. The general case has been treated in Ref. \onlinecite{BechNielsen1989}. The Ga plane of the NV center forms an angle $\theta = \cos^{-1}(\sqrt{\nicefrac{2}{3}}) = \SI{35.26}{\degree}$ with the applied stress direction (see FIG. \ref{fig:orientation_model}). From this angle $\theta$ and the specific isotope configuration, all possible angles $\theta$' are easily calculated. The resulting intensity ratios under polarized light are given in TAB \ref{tab:uniaxial_stress}. For the \textit{E} mode, theory requires a splitting into two branches, corresponding to its two eigenstates, and thus to a total lifting of its intrinsic degeneracy. The orientational degeneracy is not lifted as all orientations are equivalent with respect to the \hkl<100> axis (FIG. \ref{fig:orientation_model}). Intensity ratios are also listed in TAB \ref{tab:uniaxial_stress}.
 
 In accordance with these considerations, the required number of stress-split lines was fitted to the spectra shown in FIG. \ref{fig:spectra_under_stress}. The appropriate line intensities under polarized light were calculated from the spectrum at zero pressure, taking into account the relative intensity ratios given in TAB \ref{tab:uniaxial_stress}. Line shapes were described by a mixed gaussian-lorentzian function with an initial FWHM of \SI{0.11}{cm^{-1}}. The small stress-induced line broadening (see above) was approximated by a linear increase of the FWHM of \SI{0.133}{cm^{-1} /GPa} and a proportional decrease of intensity.

\begin{table}
\caption{\label{tab:parameters} Piezospectroscopic parameters of the components of the \SI{638}{cm^{-1}}-band as determined from the polynomial fit of the stress-induced shifts.}
\begin{ruledtabular}
\begin{tabular}{lcc}
	Mode &   \multicolumn{2}{l}{Piezospectr. parameters (\si{cm^{-1}/GPa})} \\ \hline
	E  (${\alpha}$) & A$_{\alpha}$ = +2.8 &  B$_{\alpha}$ = -0.8 \\
	E  (${\delta}$) & A$_{\delta}$ = +2.9 &  B$_{\delta}$ = -0.9 \\ \hline

	A' (${\beta}$) &  A$_{\beta 1}$ = 4.5  &  A$_{\beta 2}$ = 2.2\footnotemark[1] \\ 
	A'' (${\beta}$) &  A$_{\beta 3}$ = 1.2  &  A$_{\beta 4}$ = 
	3.9\footnotemark[1]  \\ \hline

	A' (${\gamma}$) &  A$_{\gamma 1}$ = 4.6  &  A$_{\gamma 2}$ = 2.0\footnotemark[1] \\ 
	A'' (${\gamma}$) &  A$_{\gamma 3}$ = 1.2  &  A$_{\gamma 4}$ = 3.2\footnotemark[1] \\ 
		
\end{tabular}
\end{ruledtabular}
\footnotetext[1]{initial slope for $\sigma < \SI{0.05}{GPa}$}
\end{table}

\subsection{Linear and non-linear stress dependencies}

\begin{figure}
\includegraphics[width=0.5\textwidth]{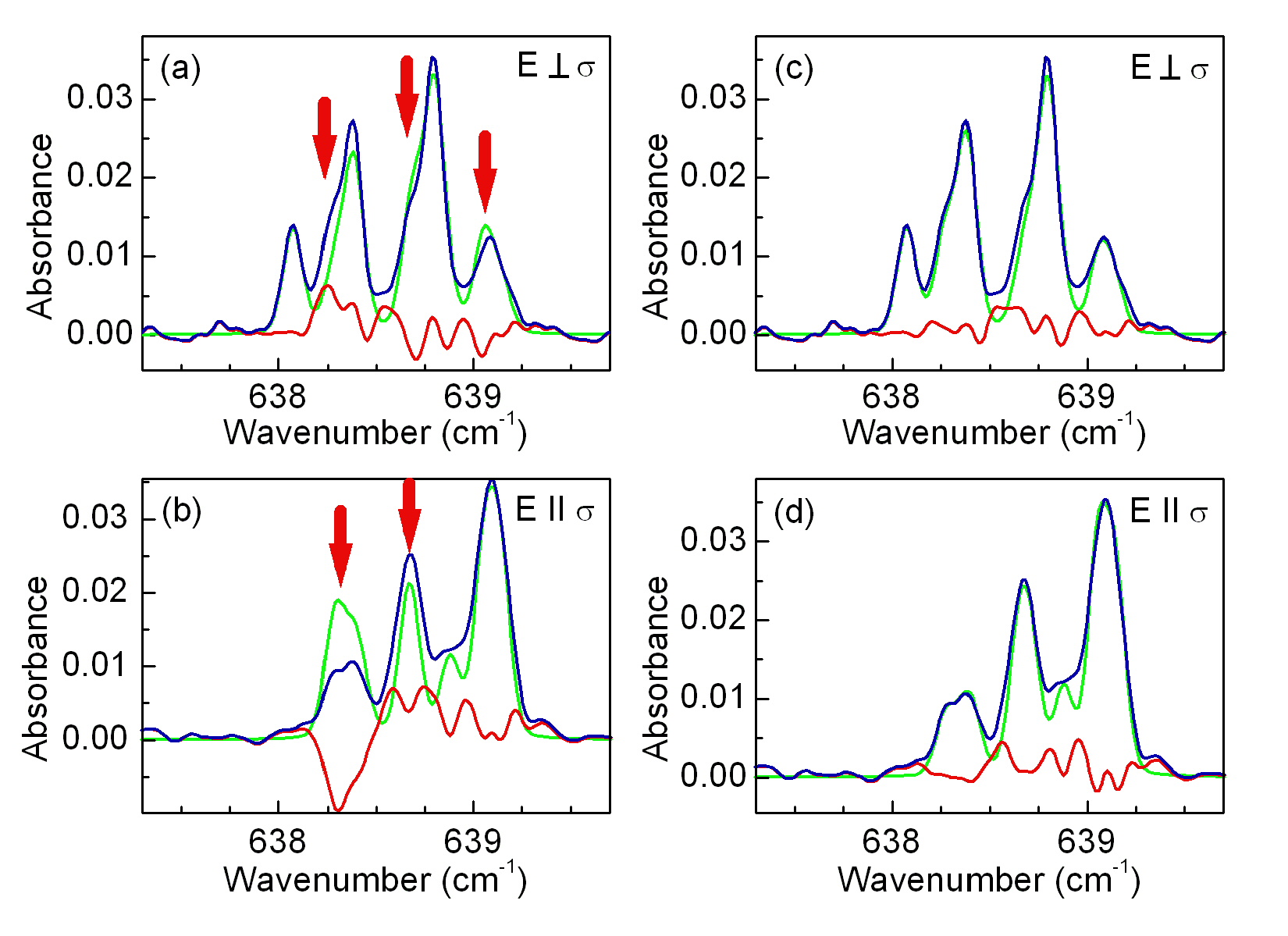}
\caption{\label{fig:fitting_illustration} Fitting results of the \SI{638}{cm^{-1}}-band at \SI{0.1}{GPa} (blue line: measured spectrum, green line: fitted curve, and  red line: difference). Linear ((a), (b)) and the improved ((c), (d)) line-shift model of the \textit{A'} and \textit{A''} modes. Significant discrepancies between the linear fit and the measured spectra are highlighted by arrows.}
\end{figure}

\begin{SCfigure*}
\includegraphics[width=0.8\textwidth]{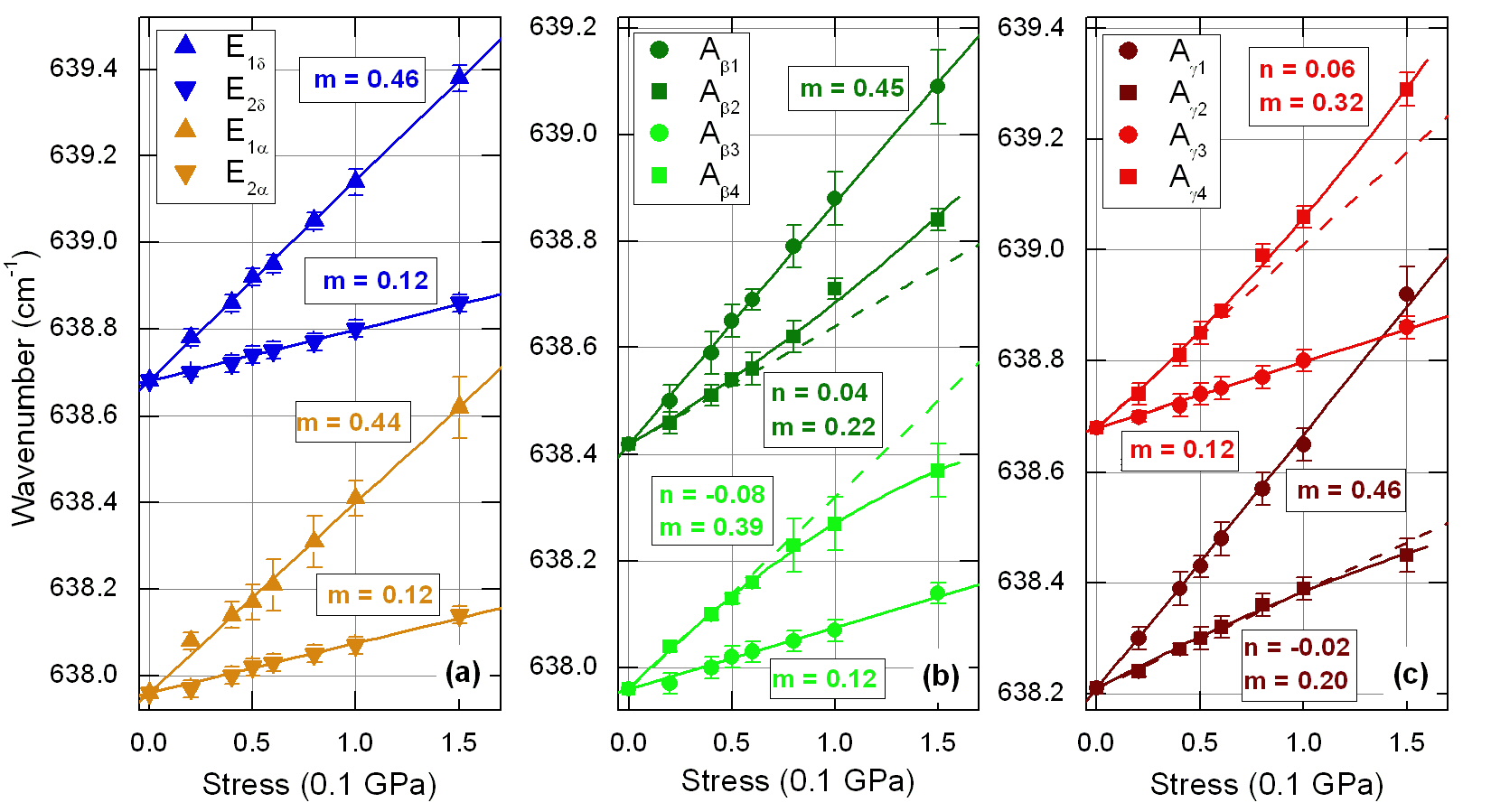}
\caption{\label{fig:final_638linefit} Shifts and splittings of all components of the \SI{638}{cm^{-1}}-band: \rb{Isotopically pure configurations are shown in (a) and mixed in (b) and (c).} The solid curves are fitted second-order polynomials, whereas the initial slopes of the non-linear branches are additionally outlined by dashed straight lines. Coefficients of linear and quadratic terms are referred to as m and n.}
\end{SCfigure*}

For low stresses, the fitted curves resulting from the linear stress dependencies listed in TAB. \ref{tab:uniaxial_stress} are in good agreement with the measured spectra. However, from \SI{0.05}{GPa} upwards, increasingly significant divergences between fitted and measured lines are obvious in some parts of the spectra (FIG. \ref{fig:fitting_illustration} (a) and (b)). In both polarization directions, however in particular for parallel polarization, lacking or excessive intensity appears within the adjusted curves. As the integrated absorbance has to be constant for all applied stresses, we must assume a stress-induced change of the intensities. Moreover, a systematic change of the stress-induced line shifts at higher stresses can be visualized from a plot of the curve fitting results so far (see FIG. \ref{fig:final_638linefit}). It shows that shifts of all lines depend linearly on the stress up to roughly \SI{0.05}{GPa}. However, starting from this stress value, four lines belonging to one branch of each \textit{A'} and \textit{A''} mode deviate from their initially linear behavior, indicating a non-linear contribution to the shift.

\begin{figure}
\includegraphics[width=0.5\textwidth]{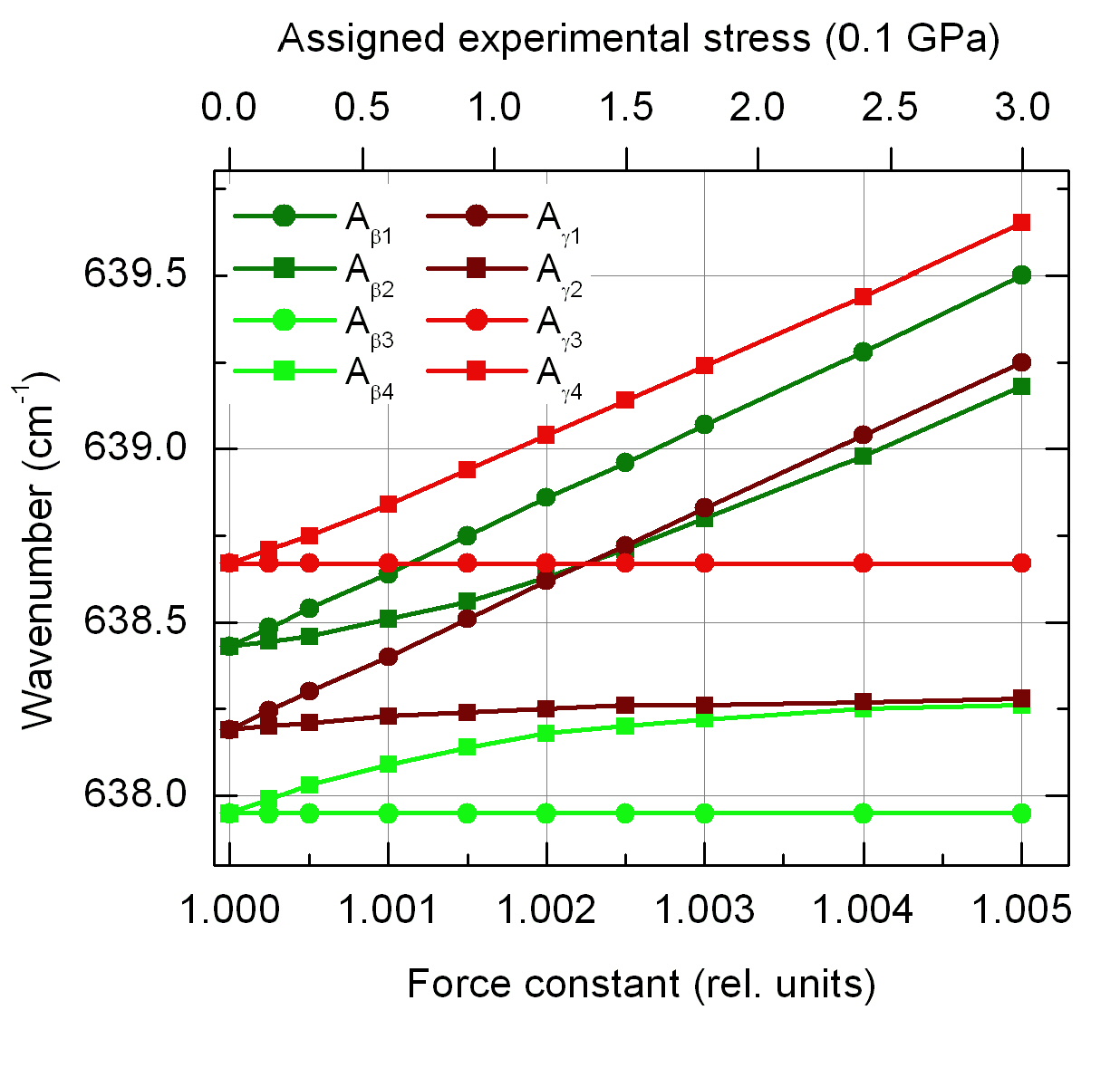}
\caption{\label{fig:calculated_nonlinearshift} Stress-induced frequency shifts of the \textit{A'} and \textit{A''} modes simulated by the valence-force model of the NV complex (circle : linear branch; square : non-linear branch).}
\end{figure}

Thus, a partial failure of the linear line-shift model is evident. Based on the observations mentioned above, these problems must be assigned to the \textit{A'} and \textit{A''} modes, originating from the mixed-isotope configurations $\beta$ and $\gamma$. The physical reason behind \rb{this} is the fact that these modes are basically \textit{E} modes which are split by two weak symmetry-lowering perturbations: (i) the deviation from \CThreeV symmetry due to the two different Ga isotopes and (ii) the deviation from \CThreeV symmetry due to the stress in \hkl<100> direction. In the present case, both effects are of the same order of magnitude. The former leads to a orientation of the two orthogonal eigenvectors ($\vec{e}_\text{H}$ and $\vec{e}_\text{L}$) of the \textit{E} mode parallel and perpendicular to the corresponding Ga isotopes (see FIG. \ref{fig:orientation_model} for one particular case). The latter tries to re-orient the two orthogonal eigenvectors such that one (with the lower frequency) is perpendicular to the stress direction. 

\begin{figure}
\includegraphics[width=0.5\textwidth]{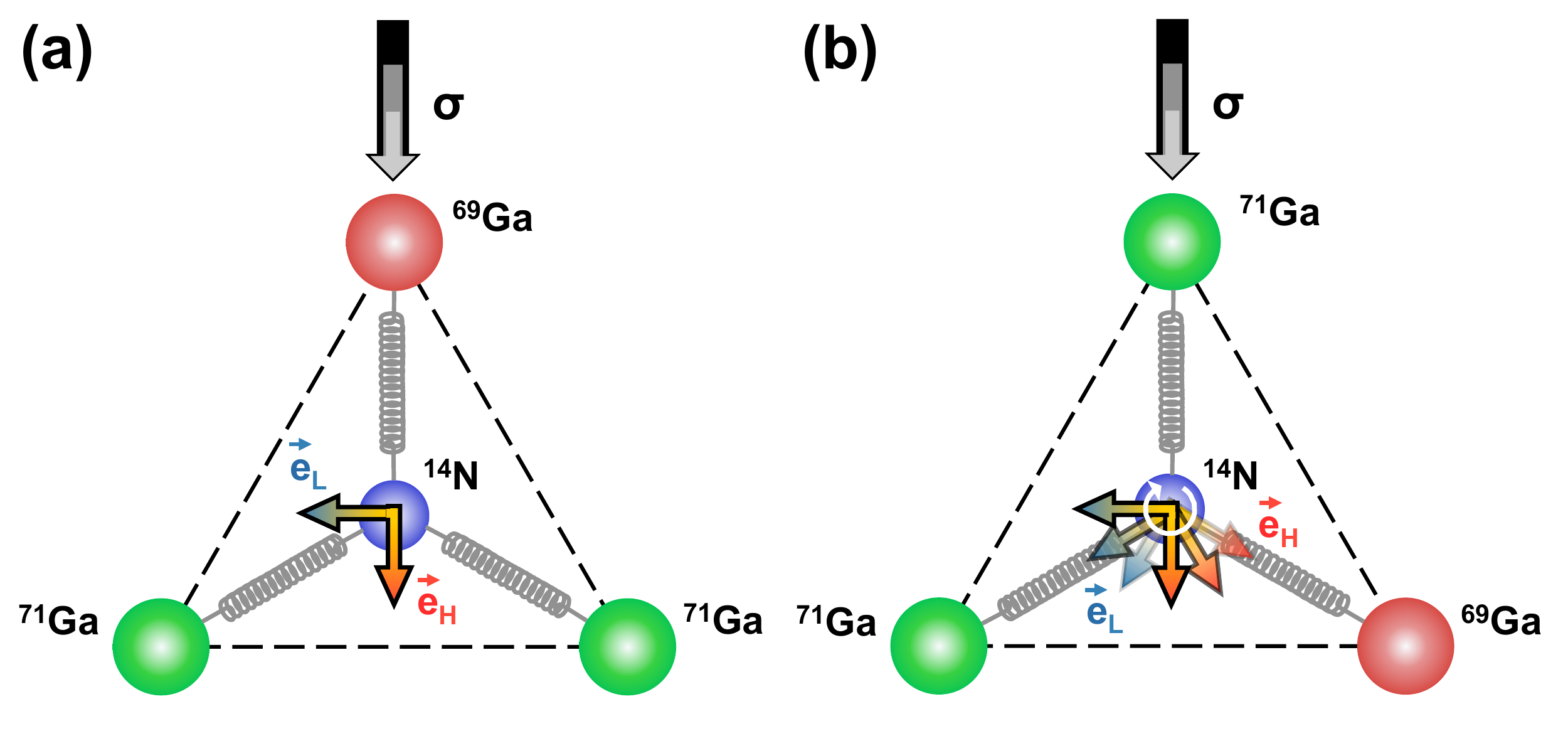}
\caption{\label{fig:eigenvector_rotation} Schematic picture of the stress-induced eigenvector rotation for the isotopically mixed NGa$_\text{3}$ configurations. The occurrence of this effect depends on the initial orientation of the high- and low-frequency eigenvectors $\vec{e}_\text{H}$ and $\vec{e}_\text{L}$ relative to the stress direction \rb{(compare (a) and (b))}.}
\end{figure}

To get more insight into the competing influences of these two perturbations, it was considered necessary to use a simulation with the empirical valence-force model (VFM) of the NV complex given earlier\cite{Alt2004}. In this model the defect is approximated by a planar NGa$_\text{3}$ molecule, orientated as shown in FIG. \ref{fig:orientation_model}. For simulating the applied stress, the force constant of the N-Ga bond including the smallest angle ($\theta$) with the \hkl<001> direction, was increased whereas the other two force constants were kept unchanged (see FIG. \ref{fig:eigenvector_rotation}). All three possibilities to distribute two \isotope[71]{Ga} and one \isotope[69]{Ga} ($\beta$) or one $\isotope[71]{Ga}$ and two $\isotope[69]{Ga}$ ($\gamma$), respectively, on the three Ga lattice sites were considered. By solving the dynamical matrix, the shift of the eigenfrequencies and the re-orientation of the eigenvectors of the vibrating system could be traced systematically as a function of stress. 

\begin{figure}
\includegraphics[width=0.5\textwidth]{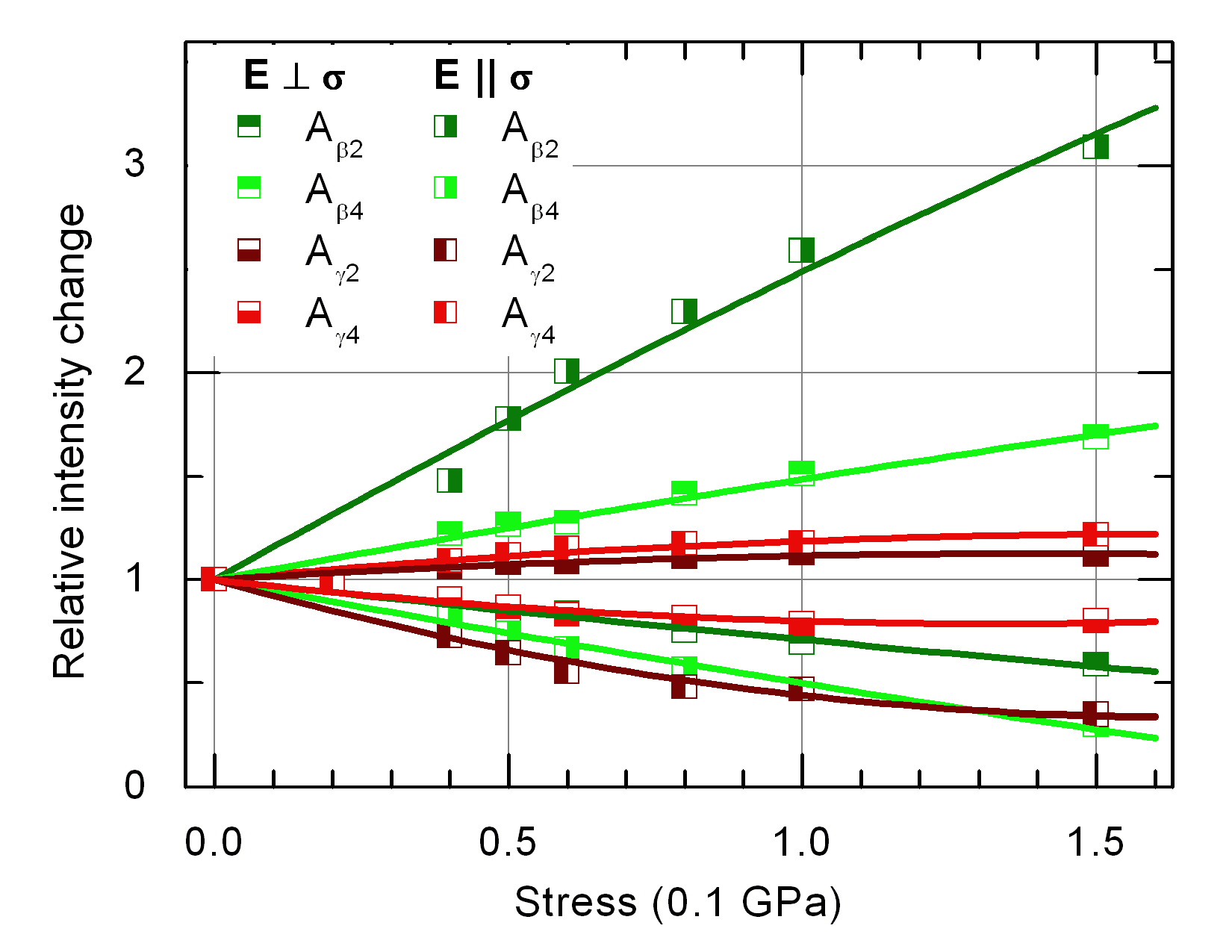}
\caption{\label{fig:fitted_intensities} Relative intensity change due to rotation of the eigenvector of the  non-linear A' and A'' branches as a function of applied stress.}
\end{figure}

The results for the eigenfrequencies are shown in FIG. \ref{fig:calculated_nonlinearshift}. In agreement with general theory\cite{Kaply1964}, the simulation predicts the splitting of each  \textit{A'} and \textit{A''} mode into two branches. In all four cases this pair consists of one branch with linear and one with non-linear stress dependence. The linear branches derive from the case where the eigenvector of the low-frequency mode is at zero stress perpendicular to the \hkl<100> direction (see FIG. \ref{fig:eigenvector_rotation} (a)), whereas the non-linear branch represent the cases where this is not true (see FIG. \ref{fig:eigenvector_rotation} (b)). In these latter cases, the non-linear behavior is due to the fact that the mode eigenvector starts rotating immediately after applying \hkl<100> stress. As this rotation is a non-linear function of the applied stress, also the frequency shift is non-linear. For both isotopically mixed configurations, the eigenvector rotation continues with increasing stress until finally the low-frequency mode is orientated perpendicular to the \hkl<100> stress axis (FIG. \ref{fig:eigenvector_rotation}). Therefore, the stress-induced shift of the non-linear branches approximates asymptotically one of the linear branches, as observable in FIG. \ref{fig:calculated_nonlinearshift}, leading finally to equal slopes.

Correlated with the mode eigenvector rotation is a change of the absorption intensity under polarized light because of the corresponding rotation of the electric dipole vector. To get an estimate of the size of this effect, we compared the simulated and measured line shifts. Thereby we could approximately assign the experimentally applied stresses values to the relative force constant changes in the VFM simulations (see FIG. \ref{fig:calculated_nonlinearshift}).The relative changes of the modified line intensities according to this approach are shown for both polarization directions in FIG. \ref{fig:fitted_intensities}. Theses corrected intensities of the four non-linear \textit{A'} and \textit{A''} modes were used to fit the spectra for stresses larger than \SI{0.04}{GPa}. Fitted spectra are now in satisfying agreement with the measured spectra (FIG. \ref{fig:fitting_illustration} (c) and (d)). Residual deviations are mostly not larger than background fluctuations. 

% \begin{SCfigure*}
% \includegraphics[width=0.7\paperwidth]{NV_Figure7_v4.pdf}
% \caption{\label{fig:final_638linefit} Shifts and splittings of all components of the \SI{638}{cm^{-1}}-band. The solid curves are fitted second-order polynomials, whereas the initial slopes of the non-linear branches are additionally outlined by dashed straight lines. Exponents of linear and quadratic terms are referred to as m and n.}
% \end{SCfigure*}

The complete results of the fitting procedure are illustrated in FIG. \ref{fig:final_638linefit}. The fitted frequency shifts are visualized by second-order polynomials. Splitting of the six modes of the NV complex into two branches each is clearly observable. The splitting originates either from total lifting of the intrinsic degeneracy in the case of the twofold degenerate \textit{E} modes, being typical for \textit{A} $\rightarrow$ \textit{E} transitions in centers of \CThreeV symmetry for \hkl<100> stress, or from partial lifting of the orientational degeneracy in the case of the \textit{A'} and \textit{A''} modes, being typical for \textit{A} $\rightarrow$ \textit{A} transitions in centers of \textit{C}$_\text{s}$ symmetry under the same conditions. 

 All branches exhibit a linear stress-induced frequency shift, apart from the four branches belonging to the \textit{A'} and \textit{A''} modes with eigenvector rotation, as discussed above. In these cases the initial (linear) slopes ($\sigma < \SI{0.05}{GPa}$) were fitted separately (see dashed lines in FIG. \ref{fig:final_638linefit}). The slopes of two \textit{E} mode branches are pairwise practically identical within the accuracy given by experiment. Indeed, it turns out that the numerical values for all slopes, when regarding the asymptotic behavior of the non-linear branches, fall into two groups: One with a slope of about \SI{1.2}{cm^{-1}/GPa} and  another with about \SI{4.6}{cm^{-1}/GPa}. This circumstance is not unexpected due to the fact that the different vibrational modes of the NV complex originate from defects being identical in structure and bonding chemistry, perturbed only slightly by the different Ga isotope combinations. Finally, all these results (splitting, shift and polarization behavior) confirm the point symmetry of the different isotope configurations for the NV complex in GaAs, given in TAB. \ref{tab:uniaxial_stress}. The corresponding piezospectroscopic parameters, derived from the linear parts of the fitted polynomials, are listed in TAB. \ref{tab:parameters}. 

\subsection{First-principles calculations}
\begin{figure}
\includegraphics[width=0.5\textwidth]{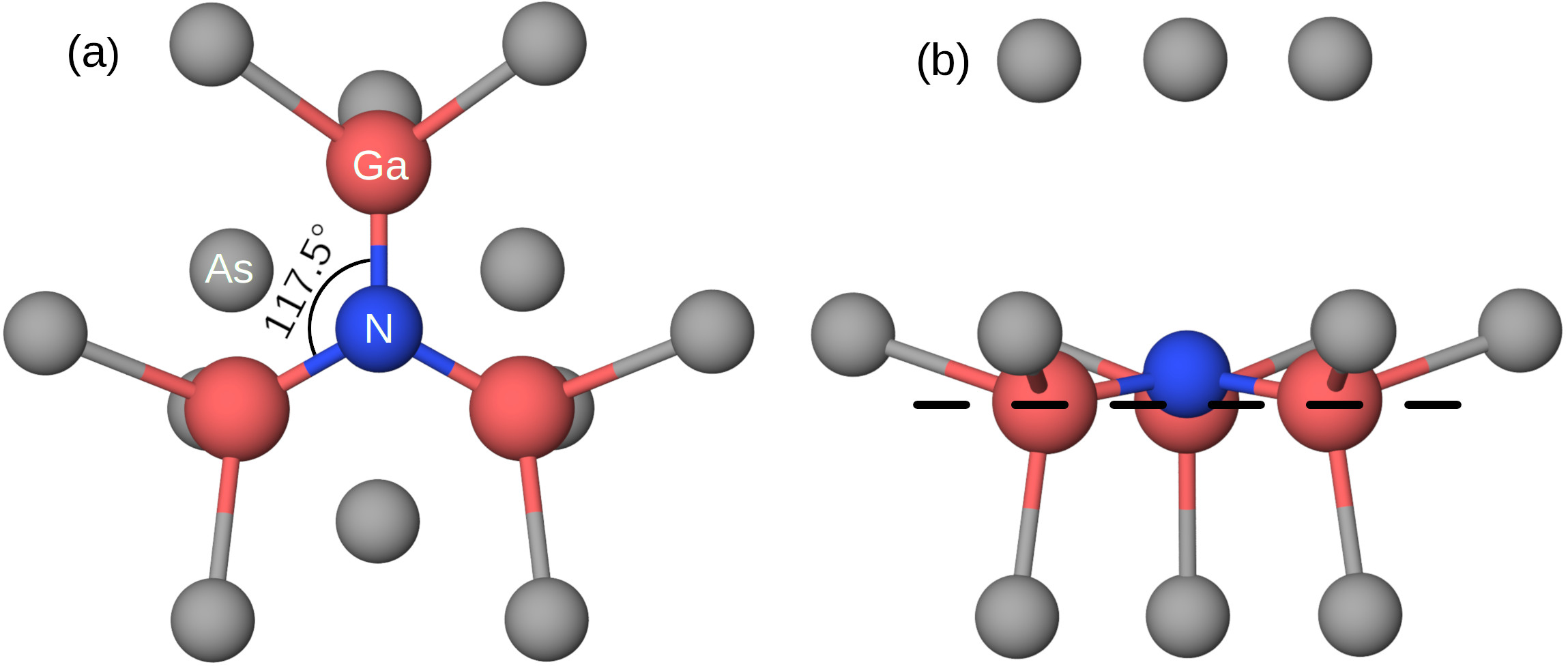}
\caption{\label{fig:structure} Local structure of the NV center in the GaAs host crystal after relaxation of the atomic positions with DFT. The N-Ga bond lengths are \SI{1.89}{\AA} (a), the distance between N and the Ga plane (dashed line) is \SI{0.76}{\AA} (b).}
\end{figure}

The starting structure for the DFT calculations was a 63-atom super cell prepared from a zinc-blende structure with 64 atoms by substituting a N atom for an As atom and removing one of the adjacent Ga atoms. The structure was converged without using symmetry constraints. The final positions in the environment of the nitrogen atom are illustrated in FIG. \ref{fig:structure}. During optimization, the N atom moves towards the center of the triangle defined by the three remaining nearest-neighbor Ga atoms. The final N-Ga bond length is \SI{1.89}{\AA} which is close to the experimental value of \SI{1.95}{\AA} in pure GaN. The relaxed site of the nitrogen atom is within \SI{0.001}{\AA} in a symmetrical position with reference to the gallium atoms, but \SI{0.76}{\AA} out of the Ga plane (see dashed line in FIG. \ref{fig:structure} (b)). Consequently, the final geometry of the NV center from DFT strongly supports the \CThreeV symmetry which has been found from the piezospectroscopic experiments. 

The vibrational frequencies for the transverse modes of the different isotopic configurations, calculated by DFT using finite displacement of all 63 atoms and measured by FTIR, turned out to be in an almost perfect agreement, thereby further confirming the interpretation of FTIR results as discussed in Section III B. The average deviation is less than \SI{2}{\%}. Scaling was performed using the formula described \rb{in} Section II. The final frequencies scaled with the factor of $\num{0.9862}$ are documented in TAB. \ref{tab:uniaxial_stress}. It should be mentioned that \rb{the} longitudinal mode frequency, with the eigenvector oriented along the direction of the nitrogen-vacancy axis, is at about \SI{310}{cm^{-1}}. This mode could not detected experimentally because is it too close to the strong optical-phonon absorption band. 

The relaxation to the minimum of energy for charged super cells leaves the position of the nitrogen atom relative to the lattice atoms unchanged within a limit of \SI{0.001}{\AA}. The result indicates that the additional or removed electrons for charging the super cell are not localized close to NV center. This could also be confirmed by analyzing the Hirshfeld charges of the DFT calculations. Furthermore, the two LVM frequencies of the N atom (tranverse and longitudinal mode) from DFT calculations vary not more than \SI{0.5}{\%} for the different charge states. 

\begin{figure}
\includegraphics[width=0.5\textwidth]{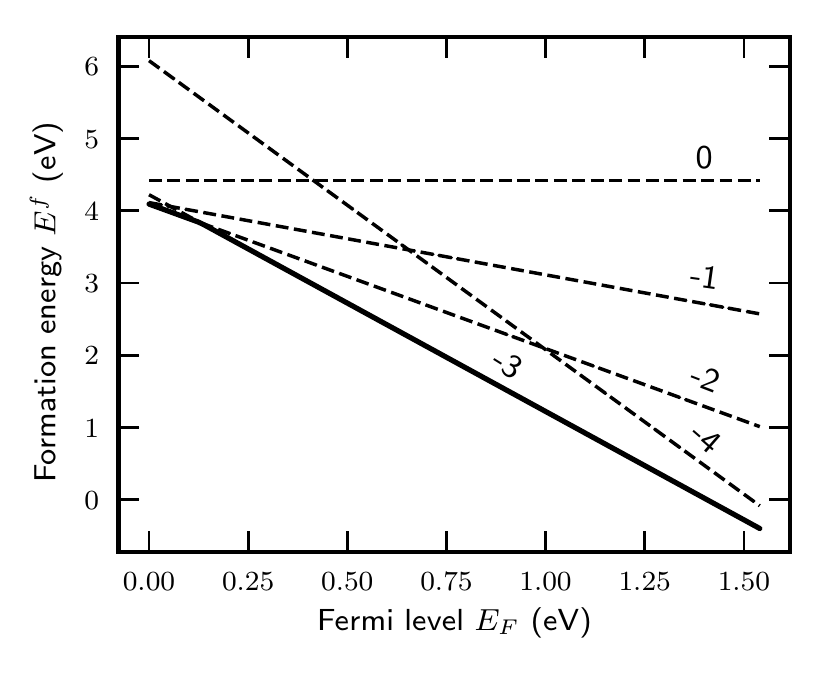}
\caption{\label{fig:formation_energy} Formation energy of the NV complex in GaAs calculated with DFT. The numbers indicate the charge state $q/e$. Positive charge states are not shown since all of them are unstable.}
\end{figure}

Formation energies, calculated by DFT using the LDA exchange-correlation functional, are shown in FIG. \ref{fig:formation_energy}. Apart from a small range up to about \SI{0.1}{eV} above the valence band edge, the $-3$ charge state has the lowest formation energy for all Fermi levels. No deep states are found in the band gap. \rb{From this calculations, the $-3$ charge state is preferred in thermodynamic equilibrium and is the charge state behind the \SI{638}{cm^{-1}}-band. However, it should be noted that the DFT calculations are carried out using LDA, which is known to underestimate the band gap and the localization of the electrons. These underestimations can affect the positions of the charge transition levels.} It appears that the $-3$ charge state is dominated by the Ga vacancy. Both positron annihilation experiments and DFT calculations find that the $-3$ charge state is prevailing for intrinsic and n-type GaAs.\cite{Gebauer2003,Mellouhi2005} The influence of the isoelectronic N atom obviously is of minor importance.

\section{Conclusion}
Piezospectroscopic FTIR studies have been carried out on the \SI{638}{cm^{-1}}-LVM of the N$_\text{As}$-V$_\text{Ga}$ nearest-neighbor pair in GaAs, the nitrogen-vacancy (NV) center. The LVM originates from the vibrational motion of the N atom perpendicular to the N$_\text{As}$-V$_\text{Ga}$ axis. The statistical occupation of the remaining three nearest-neighbor cation lattice sites with the natural occurring gallium isotopes \isotope[69]{Ga} and \isotope[71]{Ga} leads to different isotopic configurations. In isotopically pure configurations, this (transverse) LVM is twofold degenerate (\textit{E} mode) and shows for \hkl<100> stress a splitting behavior typical for \textit{A} $\rightarrow$ \textit{E} transitions in centers of \CThreeV symmetry. In the case of the mixed-isotope configurations (non-degenerate \textit{A}' and \textit{A}'' modes), the lattice deformation caused by the applied stress interacts with the perturbation due to the isotope anisotropy, because both effects are of the same order of magnitude. The resulting dynamical problem has been simulated with the help of a valence-force model. It is found that in some of the mixed-isotope configurations a significant rotation of the vibrational eigenvector and, consequently, stress-induced nonlinear frequency shifts and intensity changes have to be taken into account. However, including these effects into the analysis of the piezospectroscopic results, a complete agreement between experimentally observed and fitted absorption spectra can be achieved. 

DFT calculations on the NV complex in GaAs turned out to match the experimentally predicted \CThreeV symmetry. The N atom relaxes from the anion lattice site towards a symmetrical position with respect to the three nearest-neighbor Ga atoms. Nitrogen forms strong bonds with these Ga atoms, resulting in a triangular, nearly planar NGa$_3$ structure. The N-Ga bond length in this minimum-energy structure is close to the value in pure GaN. Furthermore, calculated formation energies show that the $-3$ charge state is lowest in energy. The NV complex has no deep defect levels in the band gap. Finally, the calculated vibrational frequency for the transverse mode is very close to the experimental value. The host isotope fine structure is reproduced perfectly. Thereby, frequency calculations further support the identification of the NV complex with the defect responsible for the \SI{638}{cm^{-1}}-LVM.  Overall, the results from FTIR experiments and DFT calculations in this work are in excellent agreement.

%\section*{Supplementary Material}
%See supplementary material for 

\begin{acknowledgments}
The authors are indebted to W. Ulrici for the supply of GaAs samples. The Gauss Centre for Supercomputing e.V. (www.gauss-centre.eu) is gratefully acknowledged for funding this project by providing computing time on the GCS Supercomputer SuperMUC at Leibniz Supercomputing Center (LRZ, www.lrz.de).
\end{acknowledgments}

\bibliography{bib}% Produces the bibliography via BibTeX.

%merlin.mbs aipnum4-1.bst 2010-07-25 4.21a (PWD, AO, DPC) hacked
%Control: key (0)
%Control: author (8) initials jnrlst
%Control: editor formatted (1) identically to author
%Control: production of article title (0) allowed
%Control: page (1) range
%Control: year (1) truncated
%Control: production of eprint (0) enabled
\begin{thebibliography}{34}%
\makeatletter
\providecommand \@ifxundefined [1]{%
 \@ifx{#1\undefined}
}%
\providecommand \@ifnum [1]{%
 \ifnum #1\expandafter \@firstoftwo
 \else \expandafter \@secondoftwo
 \fi
}%
\providecommand \@ifx [1]{%
 \ifx #1\expandafter \@firstoftwo
 \else \expandafter \@secondoftwo
 \fi
}%
\providecommand \natexlab [1]{#1}%
\providecommand \enquote  [1]{``#1''}%
\providecommand \bibnamefont  [1]{#1}%
\providecommand \bibfnamefont [1]{#1}%
\providecommand \citenamefont [1]{#1}%
\providecommand \href@noop [0]{\@secondoftwo}%
\providecommand \href [0]{\begingroup \@sanitize@url \@href}%
\providecommand \@href[1]{\@@startlink{#1}\@@href}%
\providecommand \@@href[1]{\endgroup#1\@@endlink}%
\providecommand \@sanitize@url [0]{\catcode `\\12\catcode `\$12\catcode
  `\&12\catcode `\#12\catcode `\^12\catcode `\_12\catcode `\%12\relax}%
\providecommand \@@startlink[1]{}%
\providecommand \@@endlink[0]{}%
\providecommand \url  [0]{\begingroup\@sanitize@url \@url }%
\providecommand \@url [1]{\endgroup\@href {#1}{\urlprefix }}%
\providecommand \urlprefix  [0]{URL }%
\providecommand \Eprint [0]{\href }%
\providecommand \doibase [0]{http://dx.doi.org/}%
\providecommand \selectlanguage [0]{\@gobble}%
\providecommand \bibinfo  [0]{\@secondoftwo}%
\providecommand \bibfield  [0]{\@secondoftwo}%
\providecommand \translation [1]{[#1]}%
\providecommand \BibitemOpen [0]{}%
\providecommand \bibitemStop [0]{}%
\providecommand \bibitemNoStop [0]{.\EOS\space}%
\providecommand \EOS [0]{\spacefactor3000\relax}%
\providecommand \BibitemShut  [1]{\csname bibitem#1\endcsname}%
\let\auto@bib@innerbib\@empty
%</preamble>
\bibitem [{\citenamefont {Weyers}, \citenamefont {Sato},\ and\ \citenamefont
  {Ando}(1992)}]{Weyers1992}%
  \BibitemOpen
  \bibfield  {author} {\bibinfo {author} {\bibfnamefont {M.}~\bibnamefont
  {Weyers}}, \bibinfo {author} {\bibfnamefont {M.}~\bibnamefont {Sato}}, \ and\
  \bibinfo {author} {\bibfnamefont {H.}~\bibnamefont {Ando}},\ }\bibfield
  {title} {\enquote {\bibinfo {title} {{Red shift of photoluminescence and
  absorption in dilute GaAsN alloy layers}},}\ }\href {\doibase
  10.1143/JJAP.31.L853} {\bibfield  {journal} {\bibinfo  {journal} {Jpn. J.
  Appl. Phys.}\ }\textbf {\bibinfo {volume} {31}},\ \bibinfo {pages}
  {L853--L855} (\bibinfo {year} {1992})}\BibitemShut {NoStop}%
\bibitem [{\citenamefont {Kondow}\ \emph {et~al.}(1997)\citenamefont {Kondow},
  \citenamefont {Kitatani}, \citenamefont {Nakatsuka}, \citenamefont {Larson},
  \citenamefont {Nakahara}, \citenamefont {Yazawa}, \citenamefont {Okai},\ and\
  \citenamefont {Uomi}}]{Kondow1997}%
  \BibitemOpen
  \bibfield  {author} {\bibinfo {author} {\bibfnamefont {M.}~\bibnamefont
  {Kondow}}, \bibinfo {author} {\bibfnamefont {T.}~\bibnamefont {Kitatani}},
  \bibinfo {author} {\bibfnamefont {S.}~\bibnamefont {Nakatsuka}}, \bibinfo
  {author} {\bibfnamefont {M.}~\bibnamefont {Larson}}, \bibinfo {author}
  {\bibfnamefont {K.}~\bibnamefont {Nakahara}}, \bibinfo {author}
  {\bibfnamefont {Y.}~\bibnamefont {Yazawa}}, \bibinfo {author} {\bibfnamefont
  {M.}~\bibnamefont {Okai}}, \ and\ \bibinfo {author} {\bibfnamefont
  {K.}~\bibnamefont {Uomi}},\ }\bibfield  {title} {\enquote {\bibinfo {title}
  {{GaInNAs: a novel material for long-wavelength semiconductor lasers}},}\
  }\href {\doibase 10.1109/2944.640627} {\bibfield  {journal} {\bibinfo
  {journal} {IEEE J. Sel. Top. Quantum Electron.}\ }\textbf {\bibinfo {volume}
  {3}},\ \bibinfo {pages} {719--730} (\bibinfo {year} {1997})}\BibitemShut
  {NoStop}%
\bibitem [{\citenamefont {Jackrel}\ \emph {et~al.}(2007)\citenamefont
  {Jackrel}, \citenamefont {Bank}, \citenamefont {Yuen}, \citenamefont
  {Wistey}, \citenamefont {Harris}, \citenamefont {Ptak}, \citenamefont
  {Johnston}, \citenamefont {Friedman},\ and\ \citenamefont
  {Kurtz}}]{Jackrel2007}%
  \BibitemOpen
  \bibfield  {author} {\bibinfo {author} {\bibfnamefont {D.~B.}\ \bibnamefont
  {Jackrel}}, \bibinfo {author} {\bibfnamefont {S.~R.}\ \bibnamefont {Bank}},
  \bibinfo {author} {\bibfnamefont {H.~B.}\ \bibnamefont {Yuen}}, \bibinfo
  {author} {\bibfnamefont {M.~A.}\ \bibnamefont {Wistey}}, \bibinfo {author}
  {\bibfnamefont {J.~S.}\ \bibnamefont {Harris}}, \bibinfo {author}
  {\bibfnamefont {A.~J.}\ \bibnamefont {Ptak}}, \bibinfo {author}
  {\bibfnamefont {S.~W.}\ \bibnamefont {Johnston}}, \bibinfo {author}
  {\bibfnamefont {D.~J.}\ \bibnamefont {Friedman}}, \ and\ \bibinfo {author}
  {\bibfnamefont {S.~R.}\ \bibnamefont {Kurtz}},\ }\bibfield  {title} {\enquote
  {\bibinfo {title} {{Dilute nitride GaInNAs and GaInNAsSb solar cells by
  molecular beam epitaxy}},}\ }\href {\doibase 10.1063/1.2744490} {\bibfield
  {journal} {\bibinfo  {journal} {J. Appl. Phys.}\ }\textbf {\bibinfo {volume}
  {101}},\ \bibinfo {pages} {114916} (\bibinfo {year} {2007})}\BibitemShut
  {NoStop}%
\bibitem [{\citenamefont {Wolford}\ \emph {et~al.}(1985)\citenamefont
  {Wolford}, \citenamefont {Bradley}, \citenamefont {Fry},\ and\ \citenamefont
  {Thompson}}]{Wolford1985}%
  \BibitemOpen
  \bibfield  {author} {\bibinfo {author} {\bibfnamefont {D.~J.}\ \bibnamefont
  {Wolford}}, \bibinfo {author} {\bibfnamefont {J.~A.}\ \bibnamefont
  {Bradley}}, \bibinfo {author} {\bibfnamefont {K.}~\bibnamefont {Fry}}, \ and\
  \bibinfo {author} {\bibfnamefont {J.}~\bibnamefont {Thompson}},\ }\bibfield
  {title} {\enquote {\bibinfo {title} {{The Nitrogen Isoelectronic Trap in
  GaAs}},}\ }in\ \href {\doibase 10.1007/978-1-4615-7682-2_138} {\emph
  {\bibinfo {booktitle} {Proc. 17th Int. Conf. Phys. Semicond.}}}\ (\bibinfo
  {publisher} {Springer New York},\ \bibinfo {address} {New York, NY},\
  \bibinfo {year} {1985})\ pp.\ \bibinfo {pages} {627--630}\BibitemShut
  {NoStop}%
\bibitem [{\citenamefont {Liu}\ \emph {et~al.}(1990)\citenamefont {Liu},
  \citenamefont {Pistol}, \citenamefont {Samuelson}, \citenamefont
  {Schwetlick},\ and\ \citenamefont {Seifert}}]{Liu1990}%
  \BibitemOpen
  \bibfield  {author} {\bibinfo {author} {\bibfnamefont {X.}~\bibnamefont
  {Liu}}, \bibinfo {author} {\bibfnamefont {M.-E.}\ \bibnamefont {Pistol}},
  \bibinfo {author} {\bibfnamefont {L.}~\bibnamefont {Samuelson}}, \bibinfo
  {author} {\bibfnamefont {S.}~\bibnamefont {Schwetlick}}, \ and\ \bibinfo
  {author} {\bibfnamefont {W.}~\bibnamefont {Seifert}},\ }\bibfield  {title}
  {\enquote {\bibinfo {title} {{Nitrogen pair luminescence in GaAs}},}\ }\href
  {\doibase 10.1063/1.102495} {\bibfield  {journal} {\bibinfo  {journal} {Appl.
  Phys. Lett.}\ }\textbf {\bibinfo {volume} {56}},\ \bibinfo {pages}
  {1451--1453} (\bibinfo {year} {1990})}\BibitemShut {NoStop}%
\bibitem [{\citenamefont {Francoeur}, \citenamefont {Klem},\ and\ \citenamefont
  {Mascarenhas}(2004)}]{Francoeur2004}%
  \BibitemOpen
  \bibfield  {author} {\bibinfo {author} {\bibfnamefont {S.}~\bibnamefont
  {Francoeur}}, \bibinfo {author} {\bibfnamefont {J.~F.}\ \bibnamefont {Klem}},
  \ and\ \bibinfo {author} {\bibfnamefont {A.}~\bibnamefont {Mascarenhas}},\
  }\bibfield  {title} {\enquote {\bibinfo {title} {{Optical Spectroscopy of
  Single Impurity Centers in Semiconductors}},}\ }\href {\doibase
  10.1103/PhysRevLett.93.067403} {\bibfield  {journal} {\bibinfo  {journal}
  {Phys. Rev. Lett.}\ }\textbf {\bibinfo {volume} {93}},\ \bibinfo {pages}
  {067403} (\bibinfo {year} {2004})}\BibitemShut {NoStop}%
\bibitem [{\citenamefont {Ikezawa}\ \emph {et~al.}(2012)\citenamefont
  {Ikezawa}, \citenamefont {Sakuma}, \citenamefont {Zhang}, \citenamefont
  {Sone}, \citenamefont {Mori}, \citenamefont {Hamano}, \citenamefont
  {Watanabe}, \citenamefont {Sakoda},\ and\ \citenamefont
  {Masumoto}}]{Ikezawa2012}%
  \BibitemOpen
  \bibfield  {author} {\bibinfo {author} {\bibfnamefont {M.}~\bibnamefont
  {Ikezawa}}, \bibinfo {author} {\bibfnamefont {Y.}~\bibnamefont {Sakuma}},
  \bibinfo {author} {\bibfnamefont {L.}~\bibnamefont {Zhang}}, \bibinfo
  {author} {\bibfnamefont {Y.}~\bibnamefont {Sone}}, \bibinfo {author}
  {\bibfnamefont {T.}~\bibnamefont {Mori}}, \bibinfo {author} {\bibfnamefont
  {T.}~\bibnamefont {Hamano}}, \bibinfo {author} {\bibfnamefont
  {M.}~\bibnamefont {Watanabe}}, \bibinfo {author} {\bibfnamefont
  {K.}~\bibnamefont {Sakoda}}, \ and\ \bibinfo {author} {\bibfnamefont
  {Y.}~\bibnamefont {Masumoto}},\ }\bibfield  {title} {\enquote {\bibinfo
  {title} {{Single-photon generation from a nitrogen impurity center in
  GaAs}},}\ }\href {\doibase 10.1063/1.3679181} {\bibfield  {journal} {\bibinfo
   {journal} {Appl. Phys. Lett.}\ }\textbf {\bibinfo {volume} {100}},\ \bibinfo
  {pages} {1--4} (\bibinfo {year} {2012})}\BibitemShut {NoStop}%
\bibitem [{\citenamefont {Ikezawa}\ \emph {et~al.}(2017)\citenamefont
  {Ikezawa}, \citenamefont {Zhang}, \citenamefont {Sakuma},\ and\ \citenamefont
  {Masumoto}}]{Ikezawa2017}%
  \BibitemOpen
  \bibfield  {author} {\bibinfo {author} {\bibfnamefont {M.}~\bibnamefont
  {Ikezawa}}, \bibinfo {author} {\bibfnamefont {L.}~\bibnamefont {Zhang}},
  \bibinfo {author} {\bibfnamefont {Y.}~\bibnamefont {Sakuma}}, \ and\ \bibinfo
  {author} {\bibfnamefont {Y.}~\bibnamefont {Masumoto}},\ }\bibfield  {title}
  {\enquote {\bibinfo {title} {{Quantum interference of two photons emitted
  from a luminescence center in GaAs:N}},}\ }\href {\doibase 10.1063/1.4979520}
  {\bibfield  {journal} {\bibinfo  {journal} {Appl. Phys. Lett.}\ }\textbf
  {\bibinfo {volume} {110}},\ \bibinfo {pages} {152102} (\bibinfo {year}
  {2017})}\BibitemShut {NoStop}%
\bibitem [{\citenamefont {Kent}\ and\ \citenamefont {Zunger}(2001)}]{Kent2001}%
  \BibitemOpen
  \bibfield  {author} {\bibinfo {author} {\bibfnamefont {P.~R.~C.}\
  \bibnamefont {Kent}}\ and\ \bibinfo {author} {\bibfnamefont {A.}~\bibnamefont
  {Zunger}},\ }\bibfield  {title} {\enquote {\bibinfo {title} {{Theory of
  electronic structure evolution in GaAsN and GaPN alloys}},}\ }\href {\doibase
  10.1103/PhysRevB.64.115208} {\bibfield  {journal} {\bibinfo  {journal} {Phys.
  Rev. B}\ }\textbf {\bibinfo {volume} {64}},\ \bibinfo {pages} {115208}
  (\bibinfo {year} {2001})}\BibitemShut {NoStop}%
\bibitem [{\citenamefont {Janotti}\ \emph {et~al.}(2003)\citenamefont
  {Janotti}, \citenamefont {Wei}, \citenamefont {Zhang}, \citenamefont
  {Kurtz},\ and\ \citenamefont {Van~de Walle}}]{Janotti2003}%
  \BibitemOpen
  \bibfield  {author} {\bibinfo {author} {\bibfnamefont {A.}~\bibnamefont
  {Janotti}}, \bibinfo {author} {\bibfnamefont {S.-H.}\ \bibnamefont {Wei}},
  \bibinfo {author} {\bibfnamefont {S.~B.}\ \bibnamefont {Zhang}}, \bibinfo
  {author} {\bibfnamefont {S.}~\bibnamefont {Kurtz}}, \ and\ \bibinfo {author}
  {\bibfnamefont {C.~G.}\ \bibnamefont {Van~de Walle}},\ }\bibfield  {title}
  {\enquote {\bibinfo {title} {{Interactions between nitrogen, hydrogen, and
  gallium vacancies in ${\mathrm{GaAs}}_{1\ensuremath{-}x}{\mathrm{N}}_{x}$
  alloys}},}\ }\href {\doibase 10.1103/PhysRevB.67.161201} {\bibfield
  {journal} {\bibinfo  {journal} {Phys. Rev. B}\ }\textbf {\bibinfo {volume}
  {67}},\ \bibinfo {pages} {161201} (\bibinfo {year} {2003})}\BibitemShut
  {NoStop}%
\bibitem [{\citenamefont {Spruytte}\ \emph {et~al.}(2001)\citenamefont
  {Spruytte}, \citenamefont {Coldren}, \citenamefont {Harris}, \citenamefont
  {Wampler}, \citenamefont {Krispin}, \citenamefont {Ploog},\ and\
  \citenamefont {Larson}}]{Spruytte2001}%
  \BibitemOpen
  \bibfield  {author} {\bibinfo {author} {\bibfnamefont {S.~G.}\ \bibnamefont
  {Spruytte}}, \bibinfo {author} {\bibfnamefont {C.~W.}\ \bibnamefont
  {Coldren}}, \bibinfo {author} {\bibfnamefont {J.~S.}\ \bibnamefont {Harris}},
  \bibinfo {author} {\bibfnamefont {W.}~\bibnamefont {Wampler}}, \bibinfo
  {author} {\bibfnamefont {P.}~\bibnamefont {Krispin}}, \bibinfo {author}
  {\bibfnamefont {K.}~\bibnamefont {Ploog}}, \ and\ \bibinfo {author}
  {\bibfnamefont {M.~C.}\ \bibnamefont {Larson}},\ }\bibfield  {title}
  {\enquote {\bibinfo {title} {{Incorporation of nitrogen in nitride-arsenides:
  Origin of improved luminescence efficiency after anneal}},}\ }\href {\doibase
  10.1063/1.1352675} {\bibfield  {journal} {\bibinfo  {journal} {J. Appl.
  Phys.}\ }\textbf {\bibinfo {volume} {89}},\ \bibinfo {pages} {4401--4406}
  (\bibinfo {year} {2001})}\BibitemShut {NoStop}%
\bibitem [{\citenamefont {Toivonen}\ \emph {et~al.}(2003)\citenamefont
  {Toivonen}, \citenamefont {Hakkarainen}, \citenamefont {Sopanen},
  \citenamefont {Lipsanen}, \citenamefont {Oila},\ and\ \citenamefont
  {Saarinen}}]{Toivonen2003}%
  \BibitemOpen
  \bibfield  {author} {\bibinfo {author} {\bibfnamefont {J.}~\bibnamefont
  {Toivonen}}, \bibinfo {author} {\bibfnamefont {T.}~\bibnamefont
  {Hakkarainen}}, \bibinfo {author} {\bibfnamefont {M.}~\bibnamefont
  {Sopanen}}, \bibinfo {author} {\bibfnamefont {H.}~\bibnamefont {Lipsanen}},
  \bibinfo {author} {\bibfnamefont {J.}~\bibnamefont {Oila}}, \ and\ \bibinfo
  {author} {\bibfnamefont {K.}~\bibnamefont {Saarinen}},\ }\bibfield  {title}
  {\enquote {\bibinfo {title} {{Observation of defect complexes containing Ga
  vacancies in GaAsN}},}\ }\href {\doibase 10.1063/1.1533843} {\bibfield
  {journal} {\bibinfo  {journal} {Appl. Phys. Lett.}\ }\textbf {\bibinfo
  {volume} {82}},\ \bibinfo {pages} {40--42} (\bibinfo {year}
  {2003})}\BibitemShut {NoStop}%
\bibitem [{\citenamefont {Kachare}\ \emph {et~al.}(1973)\citenamefont
  {Kachare}, \citenamefont {Spitzer}, \citenamefont {Kahan}, \citenamefont
  {Euler},\ and\ \citenamefont {Whatley}}]{Kachare1973}%
  \BibitemOpen
  \bibfield  {author} {\bibinfo {author} {\bibfnamefont {A.~H.}\ \bibnamefont
  {Kachare}}, \bibinfo {author} {\bibfnamefont {W.~G.}\ \bibnamefont
  {Spitzer}}, \bibinfo {author} {\bibfnamefont {A.}~\bibnamefont {Kahan}},
  \bibinfo {author} {\bibfnamefont {F.~K.}\ \bibnamefont {Euler}}, \ and\
  \bibinfo {author} {\bibfnamefont {T.~A.}\ \bibnamefont {Whatley}},\
  }\bibfield  {title} {\enquote {\bibinfo {title} {{Ion‐implanted nitrogen in
  gallium arsenide}},}\ }\href {\doibase 10.1063/1.1661971} {\bibfield
  {journal} {\bibinfo  {journal} {J. Appl. Phys.}\ }\textbf {\bibinfo {volume}
  {44}},\ \bibinfo {pages} {4393--4399} (\bibinfo {year} {1973})}\BibitemShut
  {NoStop}%
\bibitem [{\citenamefont {Alt}, \citenamefont {Wiedemann},\ and\ \citenamefont
  {Bethge}(1997)}]{Alt1997}%
  \BibitemOpen
  \bibfield  {author} {\bibinfo {author} {\bibfnamefont {H.}~\bibnamefont
  {Alt}}, \bibinfo {author} {\bibfnamefont {B.}~\bibnamefont {Wiedemann}}, \
  and\ \bibinfo {author} {\bibfnamefont {K.}~\bibnamefont {Bethge}},\
  }\bibfield  {title} {\enquote {\bibinfo {title} {{Spectroscopy of
  Nitrogen-Related Centers in Gallium Arsenide}},}\ }\href {\doibase
  10.4028/www.scientific.net/MSF.258-263.867} {\bibfield  {journal} {\bibinfo
  {journal} {Mater. Sci. Forum}\ }\textbf {\bibinfo {volume} {258-263}},\
  \bibinfo {pages} {867--872} (\bibinfo {year} {1997})}\BibitemShut {NoStop}%
\bibitem [{\citenamefont {G{\"{a}}rtner}\ \emph {et~al.}(1999)\citenamefont
  {G{\"{a}}rtner}, \citenamefont {Flade}, \citenamefont {Jurisch},
  \citenamefont {K{\"{o}}hler}, \citenamefont {Korb}, \citenamefont {Kretzer},\
  and\ \citenamefont {Weinert}}]{Gaertner1999}%
  \BibitemOpen
  \bibfield  {author} {\bibinfo {author} {\bibfnamefont {G.}~\bibnamefont
  {G{\"{a}}rtner}}, \bibinfo {author} {\bibfnamefont {T.}~\bibnamefont
  {Flade}}, \bibinfo {author} {\bibfnamefont {M.}~\bibnamefont {Jurisch}},
  \bibinfo {author} {\bibfnamefont {A.}~\bibnamefont {K{\"{o}}hler}}, \bibinfo
  {author} {\bibfnamefont {J.}~\bibnamefont {Korb}}, \bibinfo {author}
  {\bibfnamefont {U.}~\bibnamefont {Kretzer}}, \ and\ \bibinfo {author}
  {\bibfnamefont {B.}~\bibnamefont {Weinert}},\ }\bibfield  {title} {\enquote
  {\bibinfo {title} {{Oxygen incorporation in undoped LEC-GaAs}},}\ }\href
  {\doibase 10.1016/S0022-0248(98)01049-5} {\bibfield  {journal} {\bibinfo
  {journal} {J. Cryst. Growth}\ }\textbf {\bibinfo {volume} {198-199}},\
  \bibinfo {pages} {355--360} (\bibinfo {year} {1999})}\BibitemShut {NoStop}%
\bibitem [{\citenamefont {Alt}, \citenamefont {Gomeniuk},\ and\ \citenamefont
  {Wiedemann}(2004)}]{Alt2004}%
  \BibitemOpen
  \bibfield  {author} {\bibinfo {author} {\bibfnamefont {H.~C.}\ \bibnamefont
  {Alt}}, \bibinfo {author} {\bibfnamefont {Y.~V.}\ \bibnamefont {Gomeniuk}}, \
  and\ \bibinfo {author} {\bibfnamefont {B.}~\bibnamefont {Wiedemann}},\
  }\bibfield  {title} {\enquote {\bibinfo {title} {{Spectroscopic evidence for
  a N-Ga vacancy defect in GaAs}},}\ }\href {\doibase
  10.1103/PhysRevB.69.125214} {\bibfield  {journal} {\bibinfo  {journal} {Phys.
  Rev. B}\ }\textbf {\bibinfo {volume} {69}},\ \bibinfo {pages} {125214}
  (\bibinfo {year} {2004})}\BibitemShut {NoStop}%
\bibitem [{\citenamefont {Ulrici}, \citenamefont {Kiessling},\ and\
  \citenamefont {Rudolph}(2004)}]{Ulrici2005}%
  \BibitemOpen
  \bibfield  {author} {\bibinfo {author} {\bibfnamefont {W.}~\bibnamefont
  {Ulrici}}, \bibinfo {author} {\bibfnamefont {F.~M.}\ \bibnamefont
  {Kiessling}}, \ and\ \bibinfo {author} {\bibfnamefont {P.}~\bibnamefont
  {Rudolph}},\ }\bibfield  {title} {\enquote {\bibinfo {title} {{The
  nitrogen-hydrogen-vacancy complex in GaAs}},}\ }\href {\doibase
  10.1002/pssb.200302005} {\bibfield  {journal} {\bibinfo  {journal} {phys.
  stat. sol. (b)}\ }\textbf {\bibinfo {volume} {241}},\ \bibinfo {pages}
  {1281--1285} (\bibinfo {year} {2004})}\BibitemShut {NoStop}%
\bibitem [{\citenamefont {Blum}\ \emph {et~al.}(2009)\citenamefont {Blum},
  \citenamefont {Gehrke}, \citenamefont {Hanke}, \citenamefont {Havu},
  \citenamefont {Havu}, \citenamefont {Ren}, \citenamefont {Reuter},\ and\
  \citenamefont {Scheffler}}]{Blum2009}%
  \BibitemOpen
  \bibfield  {author} {\bibinfo {author} {\bibfnamefont {V.}~\bibnamefont
  {Blum}}, \bibinfo {author} {\bibfnamefont {R.}~\bibnamefont {Gehrke}},
  \bibinfo {author} {\bibfnamefont {F.}~\bibnamefont {Hanke}}, \bibinfo
  {author} {\bibfnamefont {P.}~\bibnamefont {Havu}}, \bibinfo {author}
  {\bibfnamefont {V.}~\bibnamefont {Havu}}, \bibinfo {author} {\bibfnamefont
  {X.}~\bibnamefont {Ren}}, \bibinfo {author} {\bibfnamefont {K.}~\bibnamefont
  {Reuter}}, \ and\ \bibinfo {author} {\bibfnamefont {M.}~\bibnamefont
  {Scheffler}},\ }\bibfield  {title} {\enquote {\bibinfo {title} {{Ab initio
  molecular simulations with numeric atom-centered orbitals}},}\ }\href
  {\doibase 10.1016/j.cpc.2009.06.022} {\bibfield  {journal} {\bibinfo
  {journal} {Comput. Phys. Commun.}\ }\textbf {\bibinfo {volume} {180}},\
  \bibinfo {pages} {2175--2196} (\bibinfo {year} {2009})}\BibitemShut {NoStop}%
\bibitem [{\citenamefont {Knuth}\ \emph {et~al.}(2015)\citenamefont {Knuth},
  \citenamefont {Carbogno}, \citenamefont {Atalla}, \citenamefont {Blum},\ and\
  \citenamefont {Scheffler}}]{Knuth2015}%
  \BibitemOpen
  \bibfield  {author} {\bibinfo {author} {\bibfnamefont {F.}~\bibnamefont
  {Knuth}}, \bibinfo {author} {\bibfnamefont {C.}~\bibnamefont {Carbogno}},
  \bibinfo {author} {\bibfnamefont {V.}~\bibnamefont {Atalla}}, \bibinfo
  {author} {\bibfnamefont {V.}~\bibnamefont {Blum}}, \ and\ \bibinfo {author}
  {\bibfnamefont {M.}~\bibnamefont {Scheffler}},\ }\bibfield  {title} {\enquote
  {\bibinfo {title} {{All-electron formalism for total energy strain
  derivatives and stress tensor components for numeric atom-centered
  orbitals}},}\ }\href {\doibase 10.1016/j.cpc.2015.01.003} {\bibfield
  {journal} {\bibinfo  {journal} {Comput. Phys. Commun.}\ }\textbf {\bibinfo
  {volume} {190}},\ \bibinfo {pages} {33--50} (\bibinfo {year}
  {2015})}\BibitemShut {NoStop}%
\bibitem [{\citenamefont {Marek}\ \emph {et~al.}(2014)\citenamefont {Marek},
  \citenamefont {Blum}, \citenamefont {Johanni}, \citenamefont {Havu},
  \citenamefont {Lang}, \citenamefont {Auckenthaler}, \citenamefont {Heinecke},
  \citenamefont {Bungartz},\ and\ \citenamefont {Lederer}}]{Marek2014}%
  \BibitemOpen
  \bibfield  {author} {\bibinfo {author} {\bibfnamefont {A.}~\bibnamefont
  {Marek}}, \bibinfo {author} {\bibfnamefont {V.}~\bibnamefont {Blum}},
  \bibinfo {author} {\bibfnamefont {R.}~\bibnamefont {Johanni}}, \bibinfo
  {author} {\bibfnamefont {V.}~\bibnamefont {Havu}}, \bibinfo {author}
  {\bibfnamefont {B.}~\bibnamefont {Lang}}, \bibinfo {author} {\bibfnamefont
  {T.}~\bibnamefont {Auckenthaler}}, \bibinfo {author} {\bibfnamefont
  {A.}~\bibnamefont {Heinecke}}, \bibinfo {author} {\bibfnamefont {H.-J.}\
  \bibnamefont {Bungartz}}, \ and\ \bibinfo {author} {\bibfnamefont
  {H.}~\bibnamefont {Lederer}},\ }\bibfield  {title} {\enquote {\bibinfo
  {title} {{The ELPA library: scalable parallel eigenvalue solutions for
  electronic structure theory and computational science}},}\ }\href {\doibase
  10.1088/0953-8984/26/21/213201} {\bibfield  {journal} {\bibinfo  {journal}
  {J. Phys. Condens. Matter}\ }\textbf {\bibinfo {volume} {26}},\ \bibinfo
  {pages} {213201} (\bibinfo {year} {2014})}\BibitemShut {NoStop}%
\bibitem [{\citenamefont {Auckenthaler}\ \emph {et~al.}(2011)\citenamefont
  {Auckenthaler}, \citenamefont {Blum}, \citenamefont {Bungartz}, \citenamefont
  {Huckle}, \citenamefont {Johanni}, \citenamefont {Kr{\"{a}}mer},
  \citenamefont {Lang}, \citenamefont {Lederer},\ and\ \citenamefont
  {Willems}}]{Auckenthaler2011}%
  \BibitemOpen
  \bibfield  {author} {\bibinfo {author} {\bibfnamefont {T.}~\bibnamefont
  {Auckenthaler}}, \bibinfo {author} {\bibfnamefont {V.}~\bibnamefont {Blum}},
  \bibinfo {author} {\bibfnamefont {H.-J.}\ \bibnamefont {Bungartz}}, \bibinfo
  {author} {\bibfnamefont {T.}~\bibnamefont {Huckle}}, \bibinfo {author}
  {\bibfnamefont {R.}~\bibnamefont {Johanni}}, \bibinfo {author} {\bibfnamefont
  {L.}~\bibnamefont {Kr{\"{a}}mer}}, \bibinfo {author} {\bibfnamefont
  {B.}~\bibnamefont {Lang}}, \bibinfo {author} {\bibfnamefont {H.}~\bibnamefont
  {Lederer}}, \ and\ \bibinfo {author} {\bibfnamefont {P.}~\bibnamefont
  {Willems}},\ }\bibfield  {title} {\enquote {\bibinfo {title} {{Parallel
  solution of partial symmetric eigenvalue problems from electronic structure
  calculations}},}\ }\href {\doibase 10.1016/j.parco.2011.05.002} {\bibfield
  {journal} {\bibinfo  {journal} {Parallel Comput.}\ }\textbf {\bibinfo
  {volume} {37}},\ \bibinfo {pages} {783--794} (\bibinfo {year}
  {2011})}\BibitemShut {NoStop}%
\bibitem [{\citenamefont {Havu}\ \emph {et~al.}(2009)\citenamefont {Havu},
  \citenamefont {Blum}, \citenamefont {Havu},\ and\ \citenamefont
  {Scheffler}}]{Havu2009}%
  \BibitemOpen
  \bibfield  {author} {\bibinfo {author} {\bibfnamefont {V.}~\bibnamefont
  {Havu}}, \bibinfo {author} {\bibfnamefont {V.}~\bibnamefont {Blum}}, \bibinfo
  {author} {\bibfnamefont {P.}~\bibnamefont {Havu}}, \ and\ \bibinfo {author}
  {\bibfnamefont {M.}~\bibnamefont {Scheffler}},\ }\bibfield  {title} {\enquote
  {\bibinfo {title} {{Efficient integration for all-electron electronic
  structure calculation using numeric basis functions}},}\ }\href {\doibase
  10.1016/j.jcp.2009.08.008} {\bibfield  {journal} {\bibinfo  {journal} {J.
  Comput. Phys.}\ }\textbf {\bibinfo {volume} {228}},\ \bibinfo {pages}
  {8367--8379} (\bibinfo {year} {2009})}\BibitemShut {NoStop}%
\bibitem [{\citenamefont {Perdew}\ and\ \citenamefont
  {Wang}(1992)}]{Perdew1992}%
  \BibitemOpen
  \bibfield  {author} {\bibinfo {author} {\bibfnamefont {J.~P.}\ \bibnamefont
  {Perdew}}\ and\ \bibinfo {author} {\bibfnamefont {Y.}~\bibnamefont {Wang}},\
  }\bibfield  {title} {\enquote {\bibinfo {title} {{Accurate and simple
  analytic representation of the electron-gas correlation energy}},}\ }\href
  {\doibase 10.1103/PhysRevB.45.13244} {\bibfield  {journal} {\bibinfo
  {journal} {Phys. Rev. B}\ }\textbf {\bibinfo {volume} {45}},\ \bibinfo
  {pages} {13244--13249} (\bibinfo {year} {1992})}\BibitemShut {NoStop}%
\bibitem [{\citenamefont {Freysoldt}\ \emph {et~al.}(2014)\citenamefont
  {Freysoldt}, \citenamefont {Grabowski}, \citenamefont {Hickel}, \citenamefont
  {Neugebauer}, \citenamefont {Kresse}, \citenamefont {Janotti},\ and\
  \citenamefont {{Van De Walle}}}]{Freysoldt2014}%
  \BibitemOpen
  \bibfield  {author} {\bibinfo {author} {\bibfnamefont {C.}~\bibnamefont
  {Freysoldt}}, \bibinfo {author} {\bibfnamefont {B.}~\bibnamefont
  {Grabowski}}, \bibinfo {author} {\bibfnamefont {T.}~\bibnamefont {Hickel}},
  \bibinfo {author} {\bibfnamefont {J.}~\bibnamefont {Neugebauer}}, \bibinfo
  {author} {\bibfnamefont {G.}~\bibnamefont {Kresse}}, \bibinfo {author}
  {\bibfnamefont {A.}~\bibnamefont {Janotti}}, \ and\ \bibinfo {author}
  {\bibfnamefont {C.~G.}\ \bibnamefont {{Van De Walle}}},\ }\bibfield  {title}
  {\enquote {\bibinfo {title} {{First-principles calculations for point defects
  in solids}},}\ }\href {\doibase 10.1103/RevModPhys.86.253} {\bibfield
  {journal} {\bibinfo  {journal} {Rev. Mod. Phys.}\ }\textbf {\bibinfo {volume}
  {86}},\ \bibinfo {pages} {253--305} (\bibinfo {year} {2014})}\BibitemShut
  {NoStop}%
\bibitem [{\citenamefont {Makov}\ and\ \citenamefont
  {Payne}(1995)}]{MakovPayne1995}%
  \BibitemOpen
  \bibfield  {author} {\bibinfo {author} {\bibfnamefont {G.}~\bibnamefont
  {Makov}}\ and\ \bibinfo {author} {\bibfnamefont {M.~C.}\ \bibnamefont
  {Payne}},\ }\bibfield  {title} {\enquote {\bibinfo {title} {{Periodic
  boundary conditions in ab initio calculations}},}\ }\href {\doibase
  10.1103/PhysRevB.51.4014} {\bibfield  {journal} {\bibinfo  {journal} {Phys.
  Rev. B}\ }\textbf {\bibinfo {volume} {51}},\ \bibinfo {pages} {4014--4022}
  (\bibinfo {year} {1995})}\BibitemShut {NoStop}%
\bibitem [{\citenamefont {Stukowski}(2009)}]{Stukowski2010}%
  \BibitemOpen
  \bibfield  {author} {\bibinfo {author} {\bibfnamefont {A.}~\bibnamefont
  {Stukowski}},\ }\bibfield  {title} {\enquote {\bibinfo {title}
  {{Visualization and analysis of atomistic simulation data with OVITO - the
  Open Visualization Tool}},}\ }\href {\doibase 10.1088/0965-0393/18/1/015012}
  {\bibfield  {journal} {\bibinfo  {journal} {Model. Simul. Mater. Sci. Eng.}\
  }\textbf {\bibinfo {volume} {18}},\ \bibinfo {pages} {015012} (\bibinfo
  {year} {2009})}\BibitemShut {NoStop}%
\bibitem [{\citenamefont {Togo}\ and\ \citenamefont {Tanaka}(2015)}]{Togo2015}%
  \BibitemOpen
  \bibfield  {author} {\bibinfo {author} {\bibfnamefont {A.}~\bibnamefont
  {Togo}}\ and\ \bibinfo {author} {\bibfnamefont {I.}~\bibnamefont {Tanaka}},\
  }\bibfield  {title} {\enquote {\bibinfo {title} {{First principles phonon
  calculations in materials science}},}\ }\href {\doibase
  10.1016/j.scriptamat.2015.07.021} {\bibfield  {journal} {\bibinfo  {journal}
  {Scr. Mater.}\ }\textbf {\bibinfo {volume} {108}},\ \bibinfo {pages} {1--5}
  (\bibinfo {year} {2015})},\ \Eprint {http://arxiv.org/abs/1506.08498}
  {1506.08498} \BibitemShut {NoStop}%
\bibitem [{\citenamefont {Irikura}, \citenamefont {Johnson},\ and\
  \citenamefont {Kacker}(2005)}]{Irikura2005}%
  \BibitemOpen
  \bibfield  {author} {\bibinfo {author} {\bibfnamefont {K.~K.}\ \bibnamefont
  {Irikura}}, \bibinfo {author} {\bibfnamefont {R.~D.}\ \bibnamefont
  {Johnson}}, \ and\ \bibinfo {author} {\bibfnamefont {R.~N.}\ \bibnamefont
  {Kacker}},\ }\bibfield  {title} {\enquote {\bibinfo {title} {{Uncertainties
  in scaling factors for ab initio vibrational frequencies}},}\ }\href
  {\doibase 10.1021/jp052793n} {\bibfield  {journal} {\bibinfo  {journal} {J.
  Phys. Chem. A}\ }\textbf {\bibinfo {volume} {109}},\ \bibinfo {pages}
  {8430--8437} (\bibinfo {year} {2005})}\BibitemShut {NoStop}%
\bibitem [{\citenamefont {Kaplyanskii}(1964)}]{Kaply1964}%
  \BibitemOpen
  \bibfield  {author} {\bibinfo {author} {\bibfnamefont {A.~A.}\ \bibnamefont
  {Kaplyanskii}},\ }\bibfield  {title} {\enquote {\bibinfo {title} {{Noncubic
  centers in cubic crystals and their piezospectroscopic investigation}},}\
  }\href@noop {} {\bibfield  {journal} {\bibinfo  {journal} {Opt. Spectrosc.}\
  }\textbf {\bibinfo {volume} {16}},\ \bibinfo {pages} {329} (\bibinfo {year}
  {1964})}\BibitemShut {NoStop}%
\bibitem [{\citenamefont {Hughes}\ and\ \citenamefont
  {Runciman}(1967)}]{Hughes1967}%
  \BibitemOpen
  \bibfield  {author} {\bibinfo {author} {\bibfnamefont {A.~E.}\ \bibnamefont
  {Hughes}}\ and\ \bibinfo {author} {\bibfnamefont {W.~A.}\ \bibnamefont
  {Runciman}},\ }\bibfield  {title} {\enquote {\bibinfo {title} {{Uniaxial
  stress splitting of doubly degenerate states of tetragonal and trigonal
  centres in cubic crystals}},}\ }\href {\doibase 10.1088/0370-1328/90/3/328}
  {\bibfield  {journal} {\bibinfo  {journal} {Proc. Phys. Soc.}\ }\textbf
  {\bibinfo {volume} {90}},\ \bibinfo {pages} {827--838} (\bibinfo {year}
  {1967})}\BibitemShut {NoStop}%
\bibitem [{\citenamefont {Davies}\ and\ \citenamefont
  {Nazare}(1980)}]{Davies1980}%
  \BibitemOpen
  \bibfield  {author} {\bibinfo {author} {\bibfnamefont {G.}~\bibnamefont
  {Davies}}\ and\ \bibinfo {author} {\bibfnamefont {M.~H.}\ \bibnamefont
  {Nazare}},\ }\bibfield  {title} {\enquote {\bibinfo {title} {{Uniaxial stress
  splitting of E to E transitions at trigonal centres in cubic crystals: the
  594 nm band in diamond}},}\ }\href {\doibase 10.1088/0022-3719/13/22/010}
  {\bibfield  {journal} {\bibinfo  {journal} {J. Phys. C Solid State Phys.}\
  }\textbf {\bibinfo {volume} {13}},\ \bibinfo {pages} {4127--4136} (\bibinfo
  {year} {1980})}\BibitemShut {NoStop}%
\bibitem [{\citenamefont {Nielsen}\ and\ \citenamefont
  {Grimmeiss}(1989)}]{BechNielsen1989}%
  \BibitemOpen
  \bibfield  {author} {\bibinfo {author} {\bibfnamefont {B.~B.}\ \bibnamefont
  {Nielsen}}\ and\ \bibinfo {author} {\bibfnamefont {H.~G.}\ \bibnamefont
  {Grimmeiss}},\ }\bibfield  {title} {\enquote {\bibinfo {title} {{Effect of
  uniaxial stress on local vibrational modes of hydrogen in ion-implanted
  silicon}},}\ }\href {\doibase 10.1103/PhysRevB.40.12403} {\bibfield
  {journal} {\bibinfo  {journal} {Phys. Rev. B}\ }\textbf {\bibinfo {volume}
  {40}},\ \bibinfo {pages} {12403--12415} (\bibinfo {year} {1989})}\BibitemShut
  {NoStop}%
\bibitem [{\citenamefont {Gebauer}\ \emph {et~al.}(2003)\citenamefont
  {Gebauer}, \citenamefont {Lausmann}, \citenamefont {Redmann}, \citenamefont
  {Krause-Rehberg}, \citenamefont {Leipner}, \citenamefont {Weber},\ and\
  \citenamefont {Ebert}}]{Gebauer2003}%
  \BibitemOpen
  \bibfield  {author} {\bibinfo {author} {\bibfnamefont {J.}~\bibnamefont
  {Gebauer}}, \bibinfo {author} {\bibfnamefont {M.}~\bibnamefont {Lausmann}},
  \bibinfo {author} {\bibfnamefont {F.}~\bibnamefont {Redmann}}, \bibinfo
  {author} {\bibfnamefont {R.}~\bibnamefont {Krause-Rehberg}}, \bibinfo
  {author} {\bibfnamefont {H.~S.}\ \bibnamefont {Leipner}}, \bibinfo {author}
  {\bibfnamefont {E.~R.}\ \bibnamefont {Weber}}, \ and\ \bibinfo {author}
  {\bibfnamefont {P.}~\bibnamefont {Ebert}},\ }\bibfield  {title} {\enquote
  {\bibinfo {title} {{Determination of the Gibbs free energy of formation of Ga
  vacancies in GaAs by positron annihilation}},}\ }\href {\doibase
  10.1103/PhysRevB.67.235207} {\bibfield  {journal} {\bibinfo  {journal} {Phys.
  Rev. B}\ }\textbf {\bibinfo {volume} {67}},\ \bibinfo {pages} {235207}
  (\bibinfo {year} {2003})}\BibitemShut {NoStop}%
\bibitem [{\citenamefont {El-Mellouhi}\ and\ \citenamefont
  {Mousseau}(2005)}]{Mellouhi2005}%
  \BibitemOpen
  \bibfield  {author} {\bibinfo {author} {\bibfnamefont {F.}~\bibnamefont
  {El-Mellouhi}}\ and\ \bibinfo {author} {\bibfnamefont {N.}~\bibnamefont
  {Mousseau}},\ }\bibfield  {title} {\enquote {\bibinfo {title}
  {{Self-vacancies inGaer gallium arsenide: An ab initio calculation}},}\
  }\href {\doibase 10.1103/PhysRevB.71.125207} {\bibfield  {journal} {\bibinfo
  {journal} {Phys. Rev. B}\ }\textbf {\bibinfo {volume} {71}},\ \bibinfo
  {pages} {125207} (\bibinfo {year} {2005})}\BibitemShut {NoStop}%
\end{thebibliography}%

\end{document}